\documentclass[twocolumn,apj,numberedappendix]{emulateapj}
\shorttitle{Unlocking Sensitivity for 21\,cm Power Spectrum Estimators}
\shortauthors{Zhang, Liu \& Parsons}
\usepackage{float}
\usepackage{amsmath}
\usepackage{graphicx}
\usepackage[figuresright]{rotating}
\usepackage{natbib}
\usepackage{gensymb}
\usepackage{ctable}
\renewcommand\[{\begin{equation}}
\renewcommand\]{\end{equation}}
\graphicspath{{figures/}}

\usepackage[breaklinks,colorlinks,citecolor=cyan]{hyperref}
\begin{document}

\title{Unlocking Sensitivity for Visibility-based Estimators of the 21\,cm Reionization Power Spectrum}

\author{
Yunfan Gerry Zhang\altaffilmark{1},
Adrian Liu\altaffilmark{1,2,3},
Aaron R. Parsons\altaffilmark{1,2}
}
\email{yunfanz@berkeley.edu}
\altaffiltext{1}{Astronomy Dept., U. California, Berkeley, CA}
\altaffiltext{2}{Radio Astronomy Lab., U. California, Berkeley, CA}
\altaffiltext{3}{Hubble Fellow}

\begin{abstract}
Radio interferometers designed to measure the cosmological 21\,cm power spectrum require high sensitivity. Several modern low-frequency interferometers feature drift-scan antennas placed on a regular grid to maximize the number of instantaneously coherent (redundant) measurements. However, even for such maximum-redundancy arrays, significant sensitivity comes through partial coherence between baselines. Current visibility-based power spectrum pipelines, though shown to ease control of systematics, lack the ability to make use of this partial redundancy. We introduce a method to leverage partial redundancy in such power spectrum pipelines for drift-scan arrays. Our method cross-multiplies baseline pairs at a time lag and quantifies the sensitivity contributions of each pair of baselines. 
Using the configurations and beams of the 128-element Donald C. Backer Precision Array for Probing the Epoch of Reionization (PAPER-128) and staged deployments of the Hydrogen Epoch of Reionization Array (HERA), we illustrate how our method applies to different arrays and predict the sensitivity improvements associated with pairing partially coherent baselines. As the number of antennas increases, we find partial redundancy to be of increasing importance in unlocking the full sensitivity of upcoming arrays. 
\end{abstract}

\section{Introduction}

The Epoch of Reionization (EoR) represents the last key
stage of our Universe's early evolution. The study of this event stands at
the intersection of cosmology and astrophysics. Understanding this
event not only serves as a scientific goal
of its own, but also as a gateway to information
regarding fundamental physics of inflation, neutrino mass, and the phenomenology
of the first stars and galaxies (e.g., \citealt{LiuOpticalDepth, Liu2016b, Mao2008, DEw21cm, Bull2015, Oyama20131186}). 

Observational studies of reionization, including Gunn-Peterson measurements of quasi-stellar objects \citep{Fan2006} and Cosmic Microwave Background temperature and anisotropy measurements (CMB; \citealt{Planck2016}), the kinetic Sunyaev-Zel'dovich effect \citep{ZahnKSZ, GeorgeKSZ, kszpatchy} and Lyman alpha galaxy observations \citep{Rhoads2017, Rhoads2012, mcquinnLyA} have given us indications of the rough time frame of reionization, but only limited constraints on the finer spatial and temporal structures. A surge of recent radio-astronomical experiments of reionization focus on measuring the ``spin-flip'' transition of neutral
hydrogen with a characteristic wavelength of 21\,cm \citep{Furlanetto2006181,PritchardLoeb}.
The 21\,cm brightness temperature is a direct tracer of neutral hydrogen through the epoch of reionization. Therefore, by measuring the three dimensional distribution of the 21\,cm signal we measure the full temporal and spatial variations of this event. 
However, before realizing full-scale 21\,cm tomography, many current radio interferometric efforts
aim to measure the spatial power spectrum of 21\,cm brightness temperature fluctuations.
Current-generation instruments include the Donald C. Backer Precision Array for Probing
the Epoch of Reionization (PAPER; \citealt{Ali2015,paper32}), the Murchison
Widefield Array (MWA; \citealt{Bowman2013, Tingay2013}), and the Low Frequency Array (LOFAR; \citealt{LOFAR}). Next-generation instruments include the Hydrogen Epoch of Reionization
Array (HERA; e.g. \citealt{HERA,JoshAntPos,HERABEAM1,HERADISH2}), which is under construction, 
and the Square Kilometer Array Low (SKA-low; e.g. \citealt{skalow2015} and references therein), which is currently in planning stages. 

The highly redshifted 21\,cm signal is faint and diffuse, in contrast to the localized bright sources targeted by many traditional radio telescopes.  Current 21\,cm experiments are sensitivity-starved, with the experimental challenge further increased when considering foreground contamination five orders of magnitude brighter than the cosmological signal of interest. Low-frequency radio interferometers aiming to measure the 21\,cm signal are thus designed differently from traditional instruments. To satisfy the sensitivity needs, modern arrays are large (ranging upwards from 100 elements) and are typically of the drift-scan (static-pointing) type to limit cost. To further improve sensitivity, experiments such as PAPER and HERA feature multiple copies of the same baselines to repeatedly measure the same Fourier signal \citep{first-paper}. 

Analysis pipelines for the 21\,cm power spectrum typically fall into two categories. In the first, images are formed through rotation synthesis, and after a foreground mitigation step, are Fourier transformed to construct a power spectrum \citep{MWAresult0, MWAresult1, LOFARresult}. An alternative technique works directly with visibilities from baselines, delay-transforming and cross-multiplying them to form the power spectrum \citep{delay-transform}. This technique avoids many systematics associated with combining data from different baselines and tracks the native sampling of the interferometer. An example of the visibility-based pipeline was presented in \cite{Ali2015}, which provided power spectrum
measurements with the 64-element version of PAPER (henceforth as PAPER-64). However, one disadvantage of existing visibility-based pipelines is their lack of treatment of partial redundancy. Baselines that are slightly different in length and orientation
``rotate into'' each other at a time offset. Baselines of different lengths and orientations therefore contain partially coherent information. This is the basis of earth-rotation synthesis. While imaging based power-spectrum pipelines naturally include all redundancy information, visibility-based pipelines so far only cross multiply fully redundant baselines, i.e., baselines of the same length and orientation.

Visibility-based pipelines are not fundamentally limited to using only fully redundant baselines. In fact, most sensitivity forecasts to date do include partial redundancy \citep{Pobersens, HERA, JoshAntPos}. A number of methods to produce power spectrum measurements using partially redundant baselines has been studied in recent years. A power-spectrum estimator based on visibilities gridded in the $uv$-plane was introduced in \cite{Choudhury2014Estimators, Choudhury2016Taperingb, Choudhury2016Taperinga}, and compared to a simple pair-baseline estimator in \cite{Choudhury2014Estimators}. Due to considerations of computational complexity, the authors favored grid-based estimator over pair-baseline estimators despite the higher accuracy of the latter.  \cite{wterm} proposed another approach based on gridded visibilities in sky-tracking measurements. 
\cite{CTrott} studied three observing strategies (tracking, drift scan, and a combination of the two) and developed a pair-visibility based approach to coherently combine visibilities gridded on the $uv$-plane for power spectrum estimation.
In this paper, we extend the above works and introduce a non-gridded \emph{baseline}-pair power spectrum estimator that can be applied to both tracking-capable arrays and drift scan-only arrays. Our formalism also provides a way to identify which baseline pairs to cross-correlate, overcoming the aforementioned computational disadvantage of baseline-based estimators relative to grid-based estimators.
Furthermore, compared to
\cite{Choudhury2014Estimators} and \cite{CTrott}, our analysis 
more explicitly treats the curved nature of the sky as well as its spectral dimension.

The basic idea of our proposed visibility-based, two-baseline power-spectrum estimator is as follows. The Earth's rotation causes the baselines in a drift-scan array to pick up different modes of the sky with time. Rotation synthesis makes use of the rotation-induced $uv$ coverage map to form an image. With visibility-based pipelines, the same information can be extracted. To do so, we cross-multiply time-shifted visibilities, with the proper weighting, to form power spectra.  Due to the large number of elements of modern arrays, the task of cross-multiplying every baseline against every other, scaling as number of array elements to the fourth power, can be computationally formidable, and many pairs of baselines provide only negligible redundancy information. Our contribution is thus twofold. First, we introduce a formalism to estimate the power spectrum from pairs of partially-redundant baselines in a visibility pipeline. Secondly we show how to use the two-baseline estimator to automatically pre-select baseline pairs and time offsets, making the problem computationally efficient. More precisely, our formalism allows one to simultaneously identify the baselines that have strong
redundancy, find the time offset that corresponds to maximal redundancy for a given pair of baselines, and quantify the sensitivity associated with cross multiplying
such a pair of partially redundant baselines, which in turn is used as weight to combine measurements in a power spectrum pipeline. 


The rest of this paper is organized as follows. In Section 2 we introduce some terminology and notation used in the rest of the paper. In Section
3 we introduce the formalism for weighting partially redundant baselines and derive the power spectrum estimator.
In Section 4 we present numerical tests of
this technique as well as the expected sensitivity improvement
 this method provides for HERA and PAPER-128 pipelines. With Section 5 we conclude. 

\section{Notation and Terminology}

In order to avoid confusion and ambiguity for the rest of this paper, we
introduce some terminology that may differ from what is commonly found in the literature. 
\begin{figure}[H]
\includegraphics[width=\linewidth]{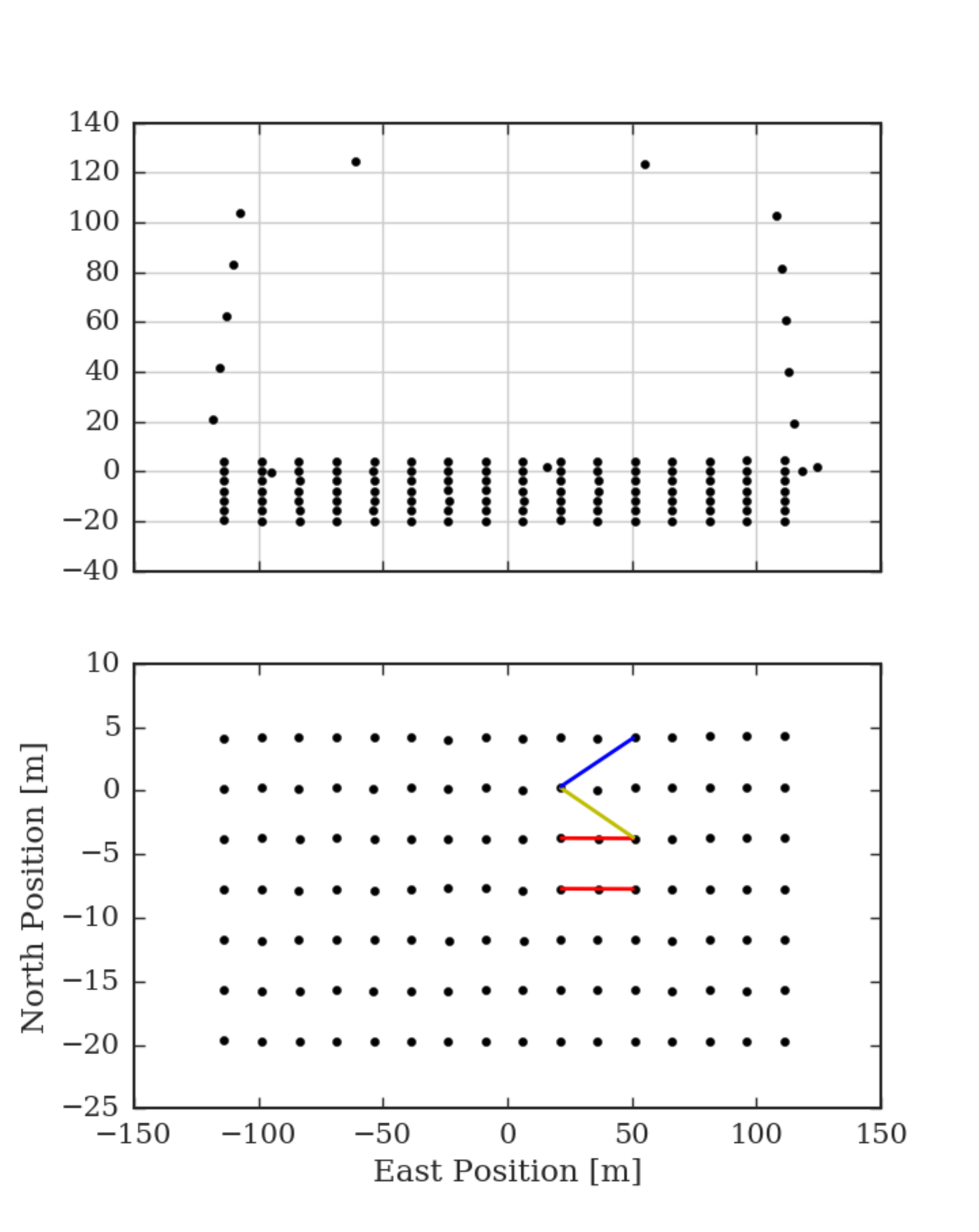}

\caption{The PAPER-128 layout. Each dot corresponds to the location of
an antenna. The top panel shows the antenna positions drawn to scale;
the bottom panel show the antenna labels and distances of the 112-element grid, excluding the outrigger antennas.
In the bottom panel, the two baselines marked with red segments are example of an equivalent pair, each with antennas separated by 2 units east and 0 unit north. Similarly,
the baselines marked in blue and yellow are examples
of classes that, due to the small North-South separation within the grid, are expected to be nearly equivalent to the red baselines. }
\label{fig:AntPos}
\end{figure}

We make the distinction between a \textit{baseline}, which corresponds to two specific antennas, and a \textit{class of baselines}, which refers to all baselines of the same length and orientation in a given array. 
Baselines of the same class are traditionally called
``redundant baselines'', because they measure the same Fourier mode
in the sky.  We shall call baselines in the same class \textit{equivalent baselines}, and reserve the \textit{redundancy} of two baselines to mean a variable function of the relative time offsets of their visibility time series. With this terminology, equivalent
baselines are fully redundant with each other simultaneously at all
times. Non-equivalent baselines also have partial redundancy, and the redundancy can be maximized 
by shifting their time series by a relative time offset.

We shall use the 128-element PAPER array (henceforth referred to as PAPER-128) to motivate our formalism and demonstrate our method, and extend our results to several HERA configurations in Section \ref{sec:arrconf}. 
The PAPER array was located in the Karoo desert in South Africa (30:43:17.5
S, 21:25:41.8 E). The layout pattern is shown
in Fig. \ref{fig:AntPos}. The array consisted of a 112-element core in a rectangular grid, and 16 ``outriggers'' used primarily to aid calibration.  In the bottom panel, the two baselines marked in red are an example of an equivalent pair. We denote a equivalency class of baselines in the PAPER grid by their separations, in this case  \{2,0\}, for the
antennas are separated by 2 units east and 0 units north. Similarly,
the baselines marked in blue and yellow are respectively examples
of \{2,1\} and $\{2, -1\}$.
Note that \{2,0\} and $\{-2, 0\}$ are the same class of baselines and should not be included twice in power-spectrum measurements. Antennas in purely north-south baselines
are close (4m), and hence these baselines are not suitable
for sensitive measurements due to cross-coupling. On the other hand, the small North-South separation means that classes such as \{2,0\} and  \{2,1\} are expected to be near-equivalent. The PAPER-64 analysis of \cite{Ali2015} used three classes of baselines, the PAPER-128
equivalent of which were
\{2, 0\}, \{2, 1\} and $\{2, -1\}$. There, baselines were cross-multiplied within each class. This paper provides the method for inter-class multiplications. We will use the short hand notation $\{m,n:m',n'\}$ to denote a pair of baseline classes to be cross-multiplied.

\section{Method \label{sec:method}}\label{sec:method}

In this section we introduce our method to cross-multiply non-equivalent baseline classes. 

\subsection{$uvw$ tracks \label{sec:tracks}}

\begin{figure}[h]
\includegraphics[width=\linewidth]{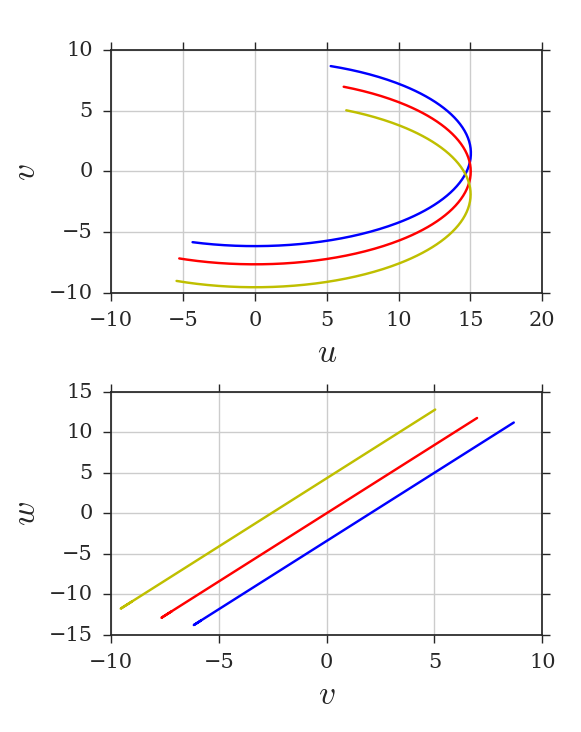}
\caption{Tracks of the three PAPER-128 baseline classes shown in Fig. \ref{fig:AntPos}, here with the same respective colors. Tracks are traced out over half a sidereal day for frequency $\nu=0.15\text{\,GHz}$ and a phase center that passes through zenith. The top panel shows $uv$ tracks with crossings among each pair of baseline classes. Projection to the $vw$ plane in the bottom panel shows that although tracks appear to overlap in the $uv$ plane, tracks in $uvw$ space do not cross. }
\label{fig:Tracks}
\end{figure}

Radio interferometric observations are often described in the coordinates $uvw$, defined as:
\[
(u, v, w) = \frac{\nu}{c}\boldsymbol{b}, 
\]
where $\boldsymbol{b}$ is the baseline vector in Cartesian coordinates, with first and second coordinates pointing East and North, respectively, and  $\nu$ is the frequency of observation, and $c$ is the speed of light. 
Relative to a phase center on the sky, each baseline maps to a point in $uvw$ space. As the Earth rotates, the points trace out tracks in the $uvw$ space. 
We show in Fig. \ref{fig:Tracks} $uvw$ tracks of the three PAPER-128 baselines colored in Fig. \ref{fig:AntPos}, projected onto the $uv$ and $vw$ planes. The tracks are traced over 12 sidereal hours, at 0.15\,GHz, relative to a phase center that drifts from zenith.

Equivalent baselines follow identical $uvw$ tracks. Traditionally, we can identify
redundancy of nearly equivalent baselines as crossings
of the $uv$ tracks, a two-dimensional projection of $uvw$, as shown in the top panel of Fig. \ref{fig:Tracks}. However, there are several reasons that $uv$ track-crossing does not imply perfect redundancy. The most obvious reason is that the three-dimensional $uvw$ tracks do not actually cross, as is evident from the $vw$ projection in the bottom panel of Fig. \ref{fig:Tracks}. In general, track-crossings are
not accurate enough to determine the optimal time offset for the cross multiplication of visibilities, nor can it provide an estimate of the degree of redundancy.   Furthermore, for drift-scan arrays, even a hypothetical crossing in $uvw$ space would not imply perfect redundancy. This will become evident in the next section, after we develop a more general formalism that accounts for the point spread function of primary. The relation between track-crossing and redundancy for both drift-scan and tracking measurements is further explored in Appendix \ref{sec:appA}.

\subsection{Formalism \label{sec:formalism}}
In this section we formulate a power spectrum estimator from
the product of delay-transformed visibilities from two arbitrary baselines. We shall work in the case of drift-scan telescopes and extend the results to tracking measurements. 

We begin with the visibility as commonly defined in the literature (e.g.
\citealt{TMS, first-paper}): 
\begin{equation}
\begin{aligned}V_{\nu}(\boldsymbol{b}) & =\int d\Omega A(\hat{\boldsymbol{s}},\nu)\phi(\nu)I_{\nu}(\hat{\boldsymbol{s}})\exp\left(-2\pi i\frac{\nu}{c}\boldsymbol{b}\cdot\hat{\boldsymbol{s}}\right)\\
 & \approx\frac{2k_{B}}{\lambda^{2}}\int d\Omega A(\hat{\boldsymbol{s}},\nu)\phi(\nu)T(\hat{\boldsymbol{s}})\exp\left(-2\pi i\frac{\nu}{c}\boldsymbol{b}\cdot\hat{\boldsymbol{s}}\right),
\end{aligned}
\label{eq:Vis1}
\end{equation}
where $k_B$ is Boltzmann's constant, $\lambda$ is the mean wavelength, and $\hat{\boldsymbol{s}}$ and $d \Omega$ denote a direction in the
sky and its corresponding solid angle. Inside the integral we have $\phi(\nu)$ as the frequency bandpass profile, $A(\hat{\boldsymbol{s}},\nu)$ as the (frequency-dependent) primary beam, and $I$ as the specific intensity, which has been
related to $T$, the brightness temperature in the Rayleigh-Jeans
limit. The beam power pattern $A$ is dimensionless,
normalized to 1 at its peak (zenith), and we assume it to be the same
for all baselines. Power-spectrum measurements are typically taken from $\sim0.01$\,GHz centered around the corresponding redshift of interest (e.g. 0.15\,GHz for redshift $z\approx9.5$), thus imposing a sharp drop-off on $\phi(\nu)$.



Ultimately we would like to relate the observed visibility to the power spectrum, $P$, defined such that
\[
\langle \widetilde{T}^{*}(\boldsymbol{k})\widetilde{T}(\boldsymbol{k'})\rangle = \left(2\pi\right)^3\delta_D^3(\bold{k}-\bold{k'})P(k),
\label{eq:Pofk}
\]
where $\bold{k}$ is the cosmological wavenumber, $\delta_D^3$ is the three-dimensional Dirac delta function, and $\widetilde{T}(\boldsymbol{k})$ is the three-dimensional Fourier transform of the brightness temperature field $T(\bold{r})$, with $\bold r$ being the cosmological position coordinate. 
We begin by relating the observational coordinates\footnote{The observational direction $\hat{\boldsymbol{s}}$ is equal to the cosmological direction $\hat{\boldsymbol{r}}$. We shall nevertheless keep the notations separate to highlight the conceptual differences and switch between the symbols depending on the context. } $\hat{\boldsymbol{s}}$
and $\nu$, to
$\boldsymbol{r}$ and $\boldsymbol{k}$: 
\[
\begin{aligned}r & =\frac{c}{H_{0}}\int_{0}^{z}\frac{dz'}{E(z')}\\
 & \approx\frac{c}{H_{0}}\int_{0}^{z_{0}}\frac{dz'}{E(z')}-\frac{c(1+z_0)^{2}}{\nu_{21}H_{0}E(z_0)}\left(\nu-\nu_{0}\right)\\
 & \equiv X-Y\Delta\nu,
\end{aligned} \label{eq:r}
\]
where $r$ is the magnitude of the vector $\bold{r}$, with $z$ its corresponding redshift. $\nu_{21}=1.420$\,GHz is the 21\,cm transition rest frequency, $\nu_{0}$ is
a reference central frequency with corresponding redshift $z_{0}$, $H_0$ is the current-day Hubble constant,  
and 
\[
E(z)=\sqrt{\Omega_{m}(1+z)^{3}+\Omega_{\Lambda}},
\]
with $\Omega_{m}$ and $\Omega_{\Lambda}$ being respectively the normalized matter and dark energy density. Inverting Eq. \eqref{eq:r} for $\nu$ gives
\begin{equation}
\nu=\frac{X-r}{Y}+\nu_{0}.\label{eq:nur}
\end{equation}
This provides us with a mapping between the observational frequency and the cosmological distance in the radial direction. For the angular directions, we have in the thin-shell limit
\begin{equation}
d^2r=X^2d\Omega. 
\label{eq:thinshell}
\end{equation}
Note that Eq. \eqref{eq:thinshell} does not require the flat-sky approximation; the angular integral is still performed over the curved sky. 

Eqs. \eqref{eq:nur} and \eqref{eq:thinshell} show that observational angular and spectral dimensions together correspond to the three dimensional volume of cosmological interest. Moreover, recall that the power spectrum in Eq. \eqref{eq:Pofk} is proportional to the product of Fourier transforms of the temperature field. Since Eq. \eqref{eq:Vis1} resembles a Fourier transform along the angular directions, it is natural to perform a further Fourier transform along the radial, or equivalently, frequency axis. The transform, called the delay transform \citep{PB2009}, has gained popularity recently since it has been shown that foregrounds are isolated in delay space.\footnote{We discuss foreground isolation in detail in Section \ref{sec:chromaticity}.} We define the delay-transformed visibility as:
\small
\begin{equation}
\begin{aligned}V(\boldsymbol{b},\tau) & \equiv \int d\nu V_{\nu}(\boldsymbol{b})\phi(\nu)\exp\left(2\pi i\nu\tau\right)\\
 & =\int d\Omega d\nu B(\hat{\boldsymbol{s}},\nu)T(\hat{\boldsymbol{s}},\nu)\exp\left[-2\pi i\nu\left(\frac{\boldsymbol{b}\cdot\hat{\boldsymbol{s}}}{c}-\tau\right)\right]. 
\end{aligned}
\label{eq:Vb1}
\end{equation}
\normalsize
Here the delay $\tau$ is the Fourier dual of $\nu$. Eq. \eqref{eq:Vb1} expresses the delay-transformed visibility as
an integral over observational coordinates $\hat{\boldsymbol{s}}$ and $\nu$. For notational simplicity we have defined the quantity

\[
B(\hat{\boldsymbol{s}},\nu) \equiv \frac{2k_B}{\lambda^2}\phi(\nu)A(\hat{\boldsymbol{s}},\nu). 
\label{eq:B}
\]

With Eqs. \eqref{eq:nur} and \eqref{eq:thinshell}, we can rewrite the delayed-transformed visibility in cosmological coordinates as 
\small
\[
\begin{aligned}V(\boldsymbol{b},\tau) & =\int\frac{d^{3}r}{X^{2}Y}B(\boldsymbol{r})T(\boldsymbol{r})\exp\left[-2\pi i\left(\frac{\boldsymbol{b}}{c}\cdot\hat{\boldsymbol{r}}-\tau\right)\nu_{r}\right],
\end{aligned}
\]
\normalsize
\!\!where $d\nu=-dr/Y$ and $d^{3}r=-X^{2}Yd\Omega d\nu$. 
We have written $\nu_{r}$ as a reminder that $\nu$ and $r$ are related
by Eq. \eqref{eq:nur}. 

Existing visibility-based power-spectrum pipelines for redundant drift-scan  arrays relate the power-spectrum to the conjugate square of the visibilities \citep{delay-transform, paper32, Ali2015}. We would like to generalize such relations by relating the power-spectrum to the product of two visibilities from two arbitrary baselines and time offsets. 
To do this, we must account for the fact that the beam pattern of a baseline shifts relative to the sky as the Earth rotates. Here we choose to fix the sky, and denote the rotated coordinates
in the topocentric frame with the three-dimensional rotation operator $\boldsymbol{\Gamma}$:
\[
\begin{aligned}V_{\psi}(\boldsymbol{b'},\tau) & = \int\frac{d^{3}r}{X^{2}Y}B(\boldsymbol{\Gamma}\boldsymbol{r})T(\boldsymbol{r}) e^{-2\pi i\left(\frac{\boldsymbol{b}}{c}\cdot\boldsymbol{\Gamma}\hat{\boldsymbol{r}}-\tau\right)\nu_{r}-i\psi_\nu},\end{aligned}
\]
where we introduced a frequency-dependent phase $\psi_{\nu}$, which to linear-order in $\nu$ one would typically pick to correspond to the re-phasing of the two visibilities to the same phase center.\footnote{For many cases, the linear-order interpretation is sufficient. We keep the general term to reserve the option to determine its exact form numerically as this information comes at no additional computational cost.} 

With implicit bounds of integrals from $-\infty$ to $\infty$, we have
\begin{eqnarray}
 &&\langle V^{*}(\boldsymbol{b},\tau)V_{\psi}(\boldsymbol{b'},\tau)\rangle \nonumber \\
 && =\int \frac{d^{3}rd^{3}r'}{(X^2Y)^2}\langle T^{*}(\boldsymbol{r})T(\boldsymbol{r'})\rangle B^{*}(\boldsymbol{r})B(\boldsymbol{\Gamma} \boldsymbol{r'})\Phi(\boldsymbol{r},\boldsymbol{\Gamma} \boldsymbol{r'}) \nonumber \\
 &&  =\int \frac{d^{3}rd^{3}r'}{(X^2Y)^2}\left(\int\frac{d^{3}\kappa}{(2\pi)^{3}}\frac{d^{3}\kappa'}{(2\pi)^{3}}\langle T^{*}(\boldsymbol{\kappa})T(\boldsymbol{\kappa'})\rangle e^{-i(\boldsymbol{\kappa}\cdot \boldsymbol{r}-\boldsymbol{\kappa'}\cdot\boldsymbol{r'})}\right) \nonumber \\
 && \qquad  \times B^{*}(\boldsymbol{r})B(\boldsymbol{\Gamma} \boldsymbol{r'})\Phi(\boldsymbol{r},\boldsymbol{\Gamma} \boldsymbol{r'})\nonumber \\
 && =\int \frac{d^{3}rd^{3}r'}{(X^2Y)^2}\left(\int\frac{d^{3}\kappa}{(2\pi)^{3}}P(\kappa)e^{-i\boldsymbol{\kappa}\cdot(\boldsymbol{r}-\boldsymbol{r'})}\right) \nonumber \\
 && \qquad \times B^{*}(\boldsymbol{r})B(\boldsymbol{\Gamma} \boldsymbol{r'})\Phi(\boldsymbol{r},\boldsymbol{\Gamma} \boldsymbol{r'}),
 \label{eq:main0}
 \end{eqnarray}
 where 
 \begin{eqnarray}
\Phi(\boldsymbol{r},\boldsymbol{\Gamma} \boldsymbol{r'})\equiv &\exp\left[i\frac{2\pi}{c}\left(\boldsymbol{b}\cdot\nu_{r}\hat{\boldsymbol{r}}-\boldsymbol{b'}\cdot\nu_{r'}\boldsymbol{\Gamma}\hat{\boldsymbol{r}}'\right)\right] \nonumber \\
& \times \exp\left[-i2\pi\tau\left(\nu_{r}-\nu_{r'}\right)-i\psi_{\nu}\right].
\end{eqnarray}
The third equality of Eq. (\ref{eq:main0}) follows from assuming the translational invariance of the statistics of the 21\,cm field. Continuing, we may make the assumption that the 3D power spectrum varies negligibly over the $k$-space of interest, thus allowing us to pull out of the integral the power spectrum centered at 
\[
k_{\boldsymbol{b}, \tau} \equiv 2\pi \sqrt{\left(\frac{\tau}{Y}\right)^2 + \left(\frac{\bar{b}}{\lambda X}\right)^2},
\label{eq:kbtau}
\]
where $\bar{b}$ is approximately the mean of $b$ and $b'$. This gives 
 \begin{eqnarray}
  &&\langle V^{*}(\boldsymbol{b},\tau)V_{\psi}(\boldsymbol{b'},\tau)\rangle \nonumber \\
 && \approx P(k_{\bar{b}, \tau})\int \frac{d^{3}rd^{3}r'}{(X^2Y)^2} \delta_{D}^{(3)}(\boldsymbol{r}-\boldsymbol{r'})B^{*}(\boldsymbol{r})B(\boldsymbol{\Gamma} \boldsymbol{r'}) \Phi(\boldsymbol{r},\boldsymbol{\Gamma} \boldsymbol{r'}) \nonumber \\
 && = P(k_{\bar{b}, \tau})\!\!\int\!\! \frac{d^{3}r}{(X^2Y)^2}B^{*}(\boldsymbol{r})B(\boldsymbol{\Gamma} \boldsymbol{r}) e^{-i2\pi\frac{\nu_{r}}{c}\left(\hat{\boldsymbol{r}}\cdot\boldsymbol{b}-\boldsymbol{\Gamma} \hat{\boldsymbol{r}}\cdot\boldsymbol{b'}\right)-i\psi_{\nu}}, \qquad
\label{eq:main}
\end{eqnarray}
where $\delta_D$ is the Dirac delta-function, and notice that the phase factor $\exp\left[-i2\pi\tau\left(\nu-\nu'\right)\right]$
drops out in the end. In factoring out the power spectrum from the integral, one is essentially making the approximation that the estimator probes only a single Fourier mode. In Section \ref{sec:chromaticity} we will relax this assumption and examine the exact form of the Fourier-space footprint that is being probed, as well as its effect on foreground isolation.

Since the beam pattern and bandpass are given in $\hat{\boldsymbol{s}}$
and $\nu$, we convert the last line of Eq. \eqref{eq:main} back to these coordinates to get
the general relation between the delay-transformed visibilities and
the power spectrum:
\begin{equation}
\begin{aligned} & \langle V^{*}(\boldsymbol{b},\tau)V_{\psi}(\boldsymbol{b'},\tau)\rangle\\
 & \approx P(k_{\bar{b}, \tau})\int\frac{d\Omega d\nu}{X^{2}Y}B^{*}(\hat{\boldsymbol{s}},\nu)B(\boldsymbol{\Gamma}\hat{\boldsymbol{s}},\nu) e^{i2\pi\frac{\nu}{c}\left(\hat{\boldsymbol{s}}\cdot\boldsymbol{b}-\boldsymbol{\Gamma}\hat{\boldsymbol{s}}\cdot\boldsymbol{b'}\right)-i\psi_{\nu}}.
 \end{aligned}
\label{eq:final}
\end{equation}
We can therefore form the power spectrum estimator for the baseline pair $\{\boldsymbol b: \boldsymbol  b'\}$:
\begin{equation}
 \hat{P}(k_{\bar{b}, \tau}) \equiv \frac{V^{*}(\boldsymbol{b},\tau)V_{\psi}(\boldsymbol{b'},\tau)}{\Theta}, 
 \label{eq:opp}
\end{equation}
where the weight is defined as 
\begin{equation}
\Theta \equiv\int d\nu \Theta_{\nu}, 
\label{eq:Theta}
\end{equation}
with
\[
\Theta_{\nu} \equiv e^{-i\psi_{\nu}} \int \frac{d\Omega}{X^{2}Y}B^{*}(\hat{\boldsymbol{s}},\nu)B(\boldsymbol{\Gamma}\hat{\boldsymbol{s}},\nu) e^{i2\pi\frac{\nu}{c}\left(\hat{\boldsymbol{s}}\cdot\boldsymbol{b}-\boldsymbol{\Gamma}\hat{\boldsymbol{s}}\cdot\boldsymbol{b'}\right)}.
\label{eq:Thetanu}
\]
Notice that $\Theta$ has no dependence on $\tau$. We point out that although all our derivations focused on drift-scan telescopes, we can get the analogous result for tracking measurements simply by noticing that for a tracking primary beam with radial symmetry we have:
\[
\Theta_{\nu} \equiv e^{-i\psi_{\nu}} \int \frac{d\Omega}{X^{2}Y}B^{*}(\hat{\boldsymbol{s}},\nu)B(\hat{\boldsymbol{s}},\nu) e^{i2\pi\frac{\nu}{c}\left(\hat{\boldsymbol{s}}\cdot\boldsymbol{b}-\boldsymbol{\Gamma}\hat{\boldsymbol{s}}\cdot\boldsymbol{b'}\right)}.
\label{eq:Thetanu_tracking}
\]

Roughly speaking, Eqs. \eqref{eq:opp} through \eqref{eq:Thetanu} tell us that the product of visibilities at a time offset is proportional to the power spectrum times the Fourier
transform of the cross multiplied beam patterns. As a check, when applied to equivalent baselines,
$\boldsymbol{b}=\boldsymbol{b'}$, the peak correlation occurs at $\Delta t=0$ and $\hat{\boldsymbol{s}}=\boldsymbol{\Gamma}\hat{\boldsymbol{s}}$, in which case Eq. \eqref{eq:final} reduces to Eq. (B9) of \cite{paper32}. 
With Eq. \eqref{eq:opp} and Eq. \eqref{eq:Theta} we can, for any given pair of baseline classes and time offsets, estimate the degree of redundancy, here represented by $\Theta$. This allows us to achieve our goals stated in the introduction: to identify 
candidate baseline pairs with significant redundancy, to find the time offset that maximizes redundancy, and to quantify the degree of such redundancy. We can do all the above simply by computing the weight $\Theta$ from
Eq. \eqref{eq:Theta} for various time offsets.

\subsection{Rephasing \label{sec:rephs}}
If $\psi_{\nu}$ were set to zero in  Eq. \eqref{eq:Thetanu}, $\Theta_{\nu}$ at the peak of correlation (as a function of time) is generally complex, and often far from real. Furthermore, this phase of peak correlation is inevitably frequency dependent. This frequency dependence would lead to destructive interference when we integrate over frequency, unless we correct $\Theta_{\nu}$ by a phase $\psi_\nu$. 

\begin{figure}[t]
\includegraphics[width=1\linewidth]{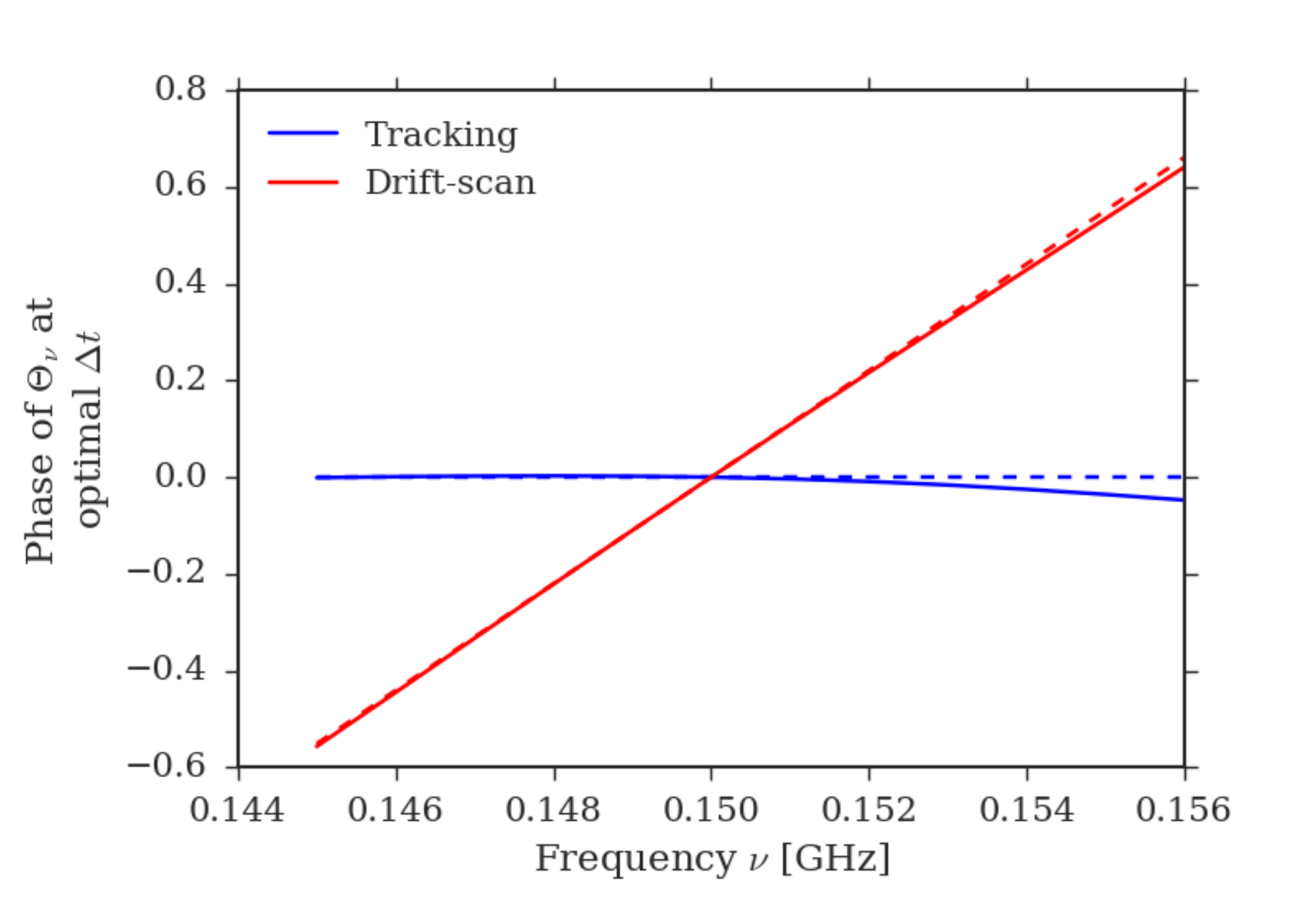}

\caption{Frequency dependent peak phases for PAPER-128 baseline pair \{2,0:2,1\}. Both the drift-scan case and a hypothetical tracking baseline are shown in solid lines. The first-order, linear effects are shown in dashed lines. We have also fixed the global phase in both cases to 0 at 0.15\,GHz. The drift-scan case exhibits linear behavior and conforms well to the first order effect of a shift in delay space due to the movement of zenith. }
\label{fig:phi_nu}
\end{figure}

Selecting $\psi_\nu$ to be a linear function of $\nu$ allows one to cancel
the decoherence due to the two visibilities having different phase centers. By default the correlators of a drift-scan array phase the two visibilities both to zenith at the same time. When they are cross-multiplied with a time lag, the visibilities must be rephased before the delay transform to account for the movement of the phase center. The effect of the phase thus roughly corresponds in a shift of the delay mode measured, and we should expect $\psi_{\nu}$ to be approximately a linear function of $\nu$:
\[
\psi_{\nu}\sim 2\pi\Delta\tau\nu,
\]
where $\Delta\tau$ is a shift in delay corresponding to the phase center movement, i.e.
\[
\Delta \tau = \frac{\boldsymbol{b'}}{c}\cdot\left(\boldsymbol{\Gamma}\hat{\boldsymbol{s}}-\hat{\boldsymbol{s}}\right). 
\]

In Fig. \ref{fig:phi_nu} we show the phases of $\Theta_\nu$ at the optimal time offset, i.e. when $|\Theta_\nu|$ is maximized. The phases are shown for the baseline pair \{2,0:2,1\} of PAPER-128. The drift-scan phase dependence is compared with that of the same baseline with hypothetical tracking-elements (Eq. \eqref{eq:Thetanu} and Eq. \eqref{eq:Thetanu_tracking}). In the drift-scan case, we indeed see a linear relation corresponding to a delay of $\Delta\tau\approx15.7$\,ns.  In the tracking case, since the phase center is fixed to the sky, only second order effects are observed. The origin of the second order effects can be seen as due to the $w$-term, or more precisely the fact that $uvw$ tracks do not cross when their two-dimensional projections do. We refer the reader to Appendix \ref{sec:appA} for further explanation. In both cases, the full effects are encapsulated in the phase of $\Theta_\nu$ and can be thus determined empirically without any additional computation. 

\begin{figure}[t]
\includegraphics[width=1.1\linewidth]{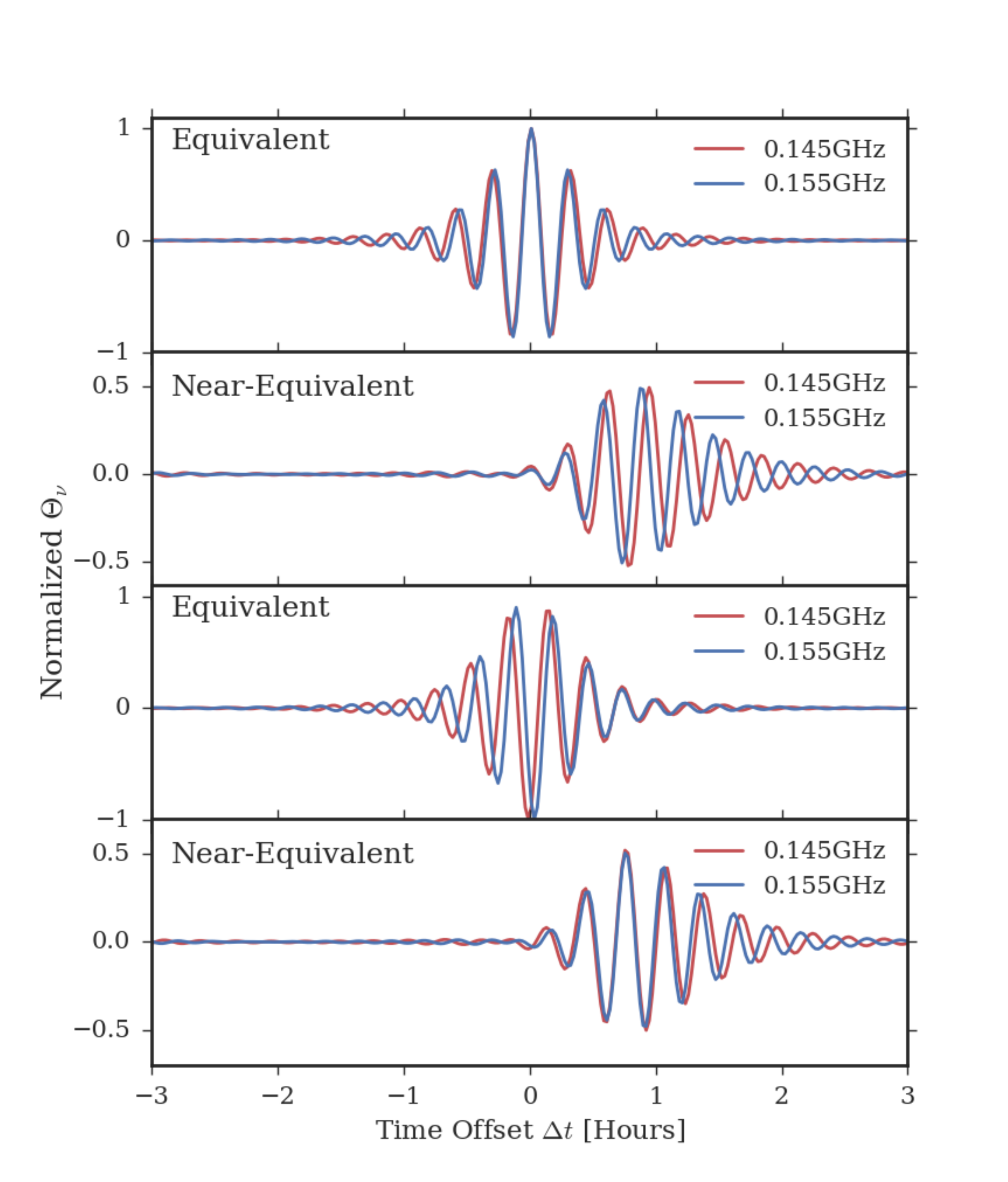}

\caption{Comparisons of the peak phases of two different frequencies. Specifically shown are the real parts of $\Theta_{\nu}$. First and third panels shows equivalent baselines, second and fourth show a pair of nearly equivalents. The top two panels have zero phase shift ($\psi_{\nu}=0$), and the bottom two are rephased to a time offset of 0.78 hours. The first and last panels thus show a coherently rephased series that would add constructively at the respective peaks. }
\label{fig:freqdiff}
\end{figure}

To illustrate the effect of rephasing, we compare in Fig. \ref{fig:freqdiff} real parts of $\Theta_{\nu}$ of both equivalent and nearly equivalent baseline pairs for two channels: 0.145\,GHz and 0.155\,GHz. The top two panels have zero rephasing, and the bottom two are linearly rephased to a differential delay of $\Delta\tau\approx15.7$\,ns, as we determined above. The first and third panels show the equivalent baseline pairs \{2,0:2,0\}, and second and fourth panels show \{2,0:2,1\}. Summing over frequency without proper rephasing leads to destructive interference and sensitivity loss. The wider the frequency profile, the more destructive the interference would be. Only after rephasing to the correct time offset for each baseline pairs can we constructively combine the frequency channels. 

We see from Fig. \ref{fig:freqdiff} that although the amplitude of correlations match up for all time offsets, the phase would only locally match. The need for rephasing can thus be understood as a symptom of the underlying spatial decoherence; coherence at the phase center does not extend to the entire beam pattern (see Section \ref{sec:visual}). While integrating over frequencies and sky-direction in Eq. \eqref{eq:Theta}, there are necessarily sky-directions that do not add coherently. This decoherence across the beam pattern cannot be removed and plays a fundamental role in determining how much sensitivity one can recover from nearly equivalent baselines. 

\section{Analysis}

In this section we delve into visualizations and analyses of our method, and explore sensitivity contributions of various baseline classes for PAPER and HERA array configurations. 
\subsection{Visualization \label{sec:visual}}

\begin{figure*}[h!]
\includegraphics[width=\textwidth]{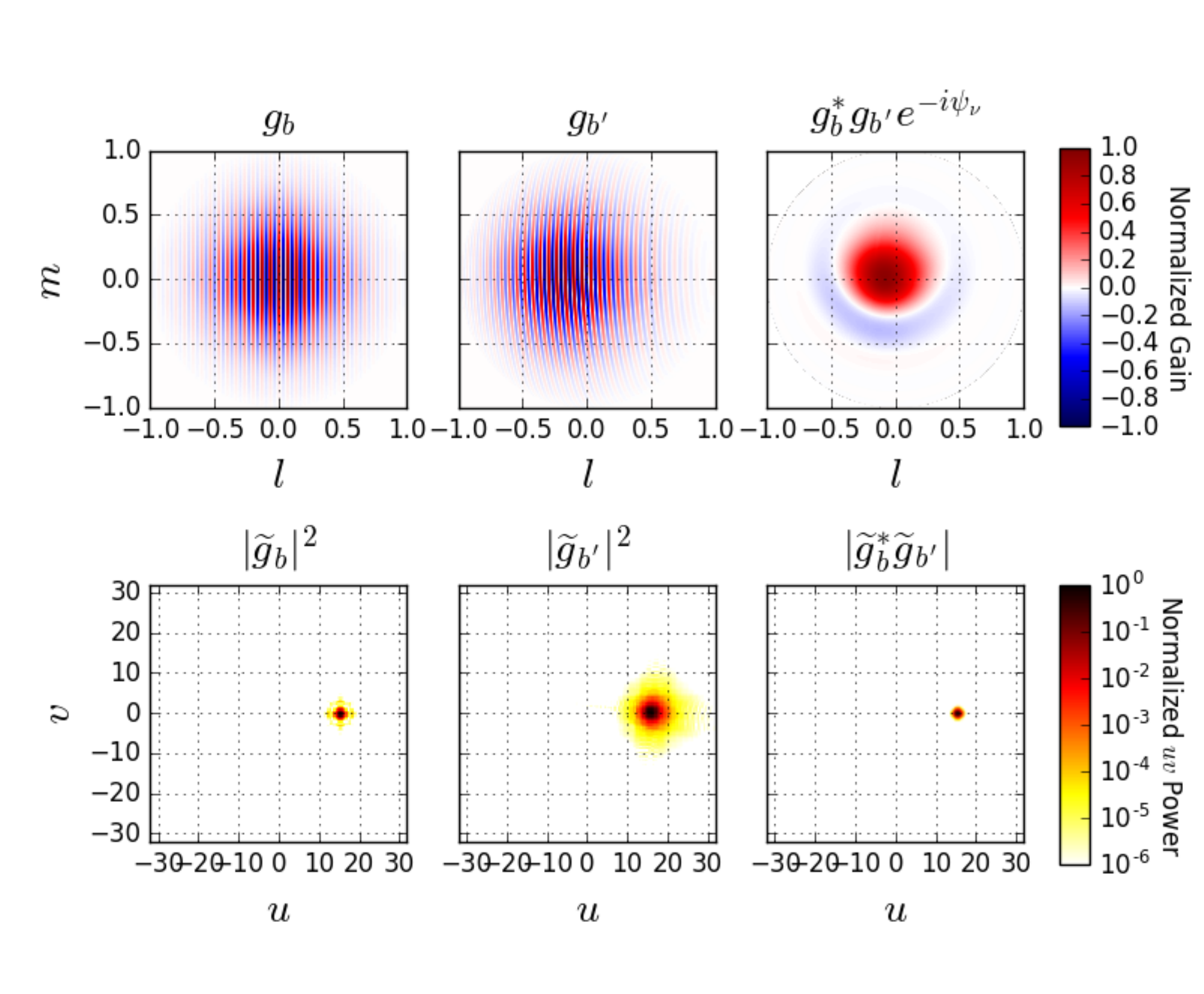}

\caption{PAPER-128 beam-fringe patterns and their point-spread functions on the $uv$-plane for a frequency of $0.15$\,GHz. Selected baselines have 30-meter East-West separations. The first baseline, shown in the first column, is purely East-West, while the second baseline, shown in the middle column, has 4-meter North-South separation and has been rotated with a time offset of 0.78 hours. Panels in the top row show the real parts of the beam-fringe patterns of the baselines (left and middle), and their conjugate product (right). The beam-fringe values are normalized such that the peak of the original beam is unity. 
The bottom-left and middle panels show the peak-normalized $uv$ point spread of the two beam-fringe patterns. The product of the beam spreads displayed in the bottom-right panel shows that power concentrated at $u\approx15$ is recovered. }
\label{fig:beamfringe}
\end{figure*}

So far for clarity and generality we have avoided the traditional formulation of radio astronomy in terms of the $uv$-plane. To gain some visual intuition of the formalism in Section \ref{sec:method}, we show explicit beam fringe patterns of the two baselines and their interactions. For this section, we let $(l,m,\sqrt{1-l^2-m^2})\equiv\hat{s}$ and examine a single frequency of $0.15$\,GHz. 

At the time offset of maximum redundancy, we expect two baselines to have to same fringe pattern in both frequency and phase. Due to the time offset, however, the beam centers would be slightly shifted with respect to each other. In the top panels of Fig. \ref{fig:beamfringe}, we show the real parts of peak-normalized beam-fringe patterns. The top left and middle panels show the beam-fringe
patterns for the baseline classes \{2,0\} and \{2,1\}, offset by 0.78 hours. More specifically, displayed are the real parts of $g_b\equiv B(\hat{\boldsymbol{s}},\nu)\exp\left[-i2\pi\frac{\nu}{c}\hat{\boldsymbol{s}}\cdot\boldsymbol{b}\right]$ and $g_{b'}\equiv B(\boldsymbol{\Gamma}\hat{\boldsymbol{s}},\nu)\exp\left[-i2\pi\frac{\nu}{c}\boldsymbol{\Gamma}\hat{\boldsymbol{s}}\cdot\boldsymbol{b'}\right]$, respectively. The top right panel show their conjugate product $g_{b}^{*}g_{b'}e^{-i\psi_\nu} = B^{*}(\hat{\boldsymbol{s}},\nu)B(\boldsymbol{\Gamma}\hat{\boldsymbol{s}},\nu)\exp\left[i2\pi\frac{\nu}{c}\left(\hat{\boldsymbol{s}}\cdot\boldsymbol{b}-\boldsymbol{\Gamma}\hat{\boldsymbol{s}}\cdot\boldsymbol{b'}\right)-i\psi_{\nu}\right]$, where we have chosen global phase $\psi_{\nu}$ such that the peak of the conjugate product is real. The product of the beam fringe patterns in the top right shows that the fringes
 cancel out as we expect. Comparing with Eq. \eqref{eq:Thetanu}, we see that $\Theta_\nu=\int d\Omega g_{b}^{*}g_{b'}e^{-i\psi_\nu}$ is the sum over the image shown in the top right panel. At the optimal time offset, the fringe patterns of the two baselines cancel out, leading to a substantial weight $\Theta_\nu$. The frequency-summed weight $\Theta$ is constructed with rephasing as described in Section \ref{sec:rephs} so that the conjugate product (top-right panel) for different frequencies align in phase at their peaks. Note it is not possible to align the phase of $g_{b}^{*}g_{b'}e^{-i\psi_\nu}$ for all $(l,m,\nu)$ due to spatial decoherence across the beam as discussed at the end of Section \ref{sec:rephs}. 
 
Similarly, we show the correlation of the two baselines in terms of the overlap of their point-spread functions in the $uv$-plane. The bottom-left and middle panels are the peak-normalized instantaneous $uv$ coverage of two baselines ($\widetilde{g}_{b}$ and $\widetilde{g}_{b'}$) while the bottom right panel is their product. We see from the bottom right panel that the cross-multiplied baselines' recovered power is concentrated at $u\approx15$. The product of the $uv$ point-spread functions shown in the bottom-right panel should not be confused with the Fourier transform of the beam-fringe conjugate product shown in the top-right panel.

\subsection{Chromaticity \label{sec:chromaticity}}

A crucial concern when measuring the $21\,\textrm{cm}$ power spectrum is the possibility of astrophysical foreground contamination. Because foregrounds are orders of magnitude larger than the cosmological signal, even small residuals from an imperfect isolation of these foregrounds can masquerade as a false detection. Recent literature treatments of this problem have shown that if one combines the mathematical properties of interferometric measurement with the empirical fact that foregrounds are spectrally smooth, a clean separation of foregrounds and cosmological signal can be achieved in harmonic space. These studies suggest that foregrounds preferentially appear in harmonic modes of the sky corresponding to fluctuations that are angularly fine but spectrally smooth, a region in harmonic space colloquially known as ``the wedge" \citep{Datta2010,Vedantham2012,Morales2012,delay-transform,Trott2012,Thyagarajan2013,pober_et_al2013b,dillon_et_al2014,Hazelton2013,Thyagarajan_et_al2015a,Thyagarajan_et_al2015b,wedge1, wedge2,chapman_et_al2016,pober_et_al2016,seo_and_hirata2016,jensen_et_al2016,kohn_et_al2016}. A foreground mitigation strategy can then consist mainly of avoiding those modes by working in the complementary region known as ``the EoR window". However, for this strategy to be viable, it is necessary to show that one's data analysis pipeline does not corrupt the separation between the wedge and the EoR window. We will now do so for our proposed two-baseline power spectrum estimator.


Traditionally, spectral and angular dimensions are cylindrically binned with the coordinates $k_\bot$ and  $k_\parallel$, where the angular direction $k_\bot$ is probed by the beam and line-of-sight direction $k_\parallel$ is probed by the frequency spectrum. Despite their usefulness, however, $(k_\bot, k_\parallel)$ are local coordinates that are well defined only in the narrow field-of-view limit. 
As discussed in previous sections, the combination of information from nearly equivalent baselines for wide-field telescopes is inherently a three-dimensional problem. Although the estimator of Section \ref{sec:method} does not have such limitations, to discuss foreground signatures with possible curved sky effects, we must employ a set of basis functions where angular fluctuations are encoded by spherical harmonics, and radial fluctuations by spherical Bessel functions. 

\cite{sphere} discusses in detail power-spectrum analyses using spherical harmonics. In such a basis, modes in the sky are indexed by the magnitude $k$ of the wavevector $\mathbf{k}$ and spherical harmonic indices $(\ell, m)$. In other words, one defines
\begin{equation}
\overline{T}_{\ell m} (k) \equiv \sqrt{\frac{2}{\pi}} \int d\Omega dr r^2 j_\ell (kr) Y_{\ell m}^* (\hat{\mathbf{r}}) T (\hat{\mathbf{r}}),
\end{equation}
where $Y_{\ell m}$ is a spherical harmonic function and $j_\ell$ is the $\ell$th-order spherical Bessel function of the first kind. The power spectrum is then related to these spherical harmonic Bessel modes via the relation
\begin{equation}
\label{eq:SHB}
\langle \overline{T}_{\ell m} (k) \overline{T}_{\ell^\prime m^\prime}^* (k^\prime) \rangle = \frac{\delta^D (k - k^\prime)}{k^2} \delta_{\ell \ell^\prime} \delta_{m m^\prime} P(k).
\end{equation}

A delay-transformed visibility $V(\mathbf{b}, \tau) $ is related to the spherical Fourier-Bessel modes by
\begin{equation}
V(\mathbf{b}, \tau) = \sqrt{\frac{2}{\pi}} \sum_{\ell m} \int \! dk\, k^2 g_{\ell m} (k; \mathbf{b}, \tau) \overline{T}_{\ell m} (k),
\end{equation}
where
\begin{equation}
g_{\ell m} (k ; \mathbf{b}, \tau) = \int \! d\Omega d\nu B(\hat{\mathbf{r}},\nu) Y_{\ell m} (\hat{\mathbf{r}}) j_\ell (k r) e^{i 2 \pi \nu (\tau - \mathbf{b}\cdot \hat{\mathbf{r}} / c)}
\end{equation}
and the analogous quantity for $V_\psi$ is given by
\begin{eqnarray}
\label{eq:glmpsi}
g_{\ell m}^\psi (k ; \mathbf{b}, \tau) = \int && d\Omega d\nu B(\boldsymbol \Gamma \hat{\mathbf{r}},\nu) Y_{\ell m} (\hat{\mathbf{r}}) j_\ell (k r) \nonumber \\
&& \times  e^{i 2 \pi \nu (\tau - \mathbf{b}\cdot \boldsymbol \Gamma \hat{\mathbf{r}} / c) + i \psi_\nu}.
\end{eqnarray}
Note that unlike the primary beam and the fringe pattern, the spherical harmonic function does \emph{not} rotate, since it originated from a spherical harmonic expansion of the sky temperature, which is fixed. With these expressions and a little algebra combining Eqs. \eqref{eq:SHB} through \eqref{eq:glmpsi}, we arrive at a form of the two-baseline estimator (compare with Eq. \eqref{eq:final}):
\begin{eqnarray}
\hat{P}(k_{\bar{b}, \tau}) &=& \frac{ \langle V^{*}(\boldsymbol{b},\tau)V_{\psi}(\boldsymbol{b'},\tau) \rangle}{\Theta} \nonumber \\
&=& \sum_\ell \int \!dk\, W_{\ell}(k; \mathbf{b}, \mathbf{b}^\prime , \tau) P(k),
\end{eqnarray}
where we have defined the \emph{window function}
\begin{equation}
W_{\ell}(k; \mathbf{b}, \mathbf{b}^\prime , \tau) \equiv \frac{2 k^2}{\pi \Theta} \sum_m g_{\ell m} (k ; \mathbf{b}, \tau) g_{\ell m}^{\psi*} (k ; \mathbf{b}^\prime, \tau).
\end{equation}
By construction, the window function sums to unity when integrated over all $k$ and summed over all $\ell$. We may therefore interpret our estimator $\hat{P}$ of the power spectrum to be a weighted average over all possible $\ell$ and $k$ modes, with the window function providing the weights of this average. This means that window functions can serve as indicators of foreground leakage into the EoR window. One writes down the estimator for a hypothetical measurement of a power spectrum mode within the EoR window, and additionally computes the window function for that estimator. Ideally, the window function for $\hat{P}(k)$ will be sharply peaked around $k$. In general, however, window functions will have wings that encroach upon other modes. If the window function is substantially non-zero at low-$k$ modes---which is where foreground emission typically resides, given its spectral smoothness---then the EoR window will be contaminated by foregrounds.

\begin{figure}[h]
\includegraphics[width=1.1\linewidth]{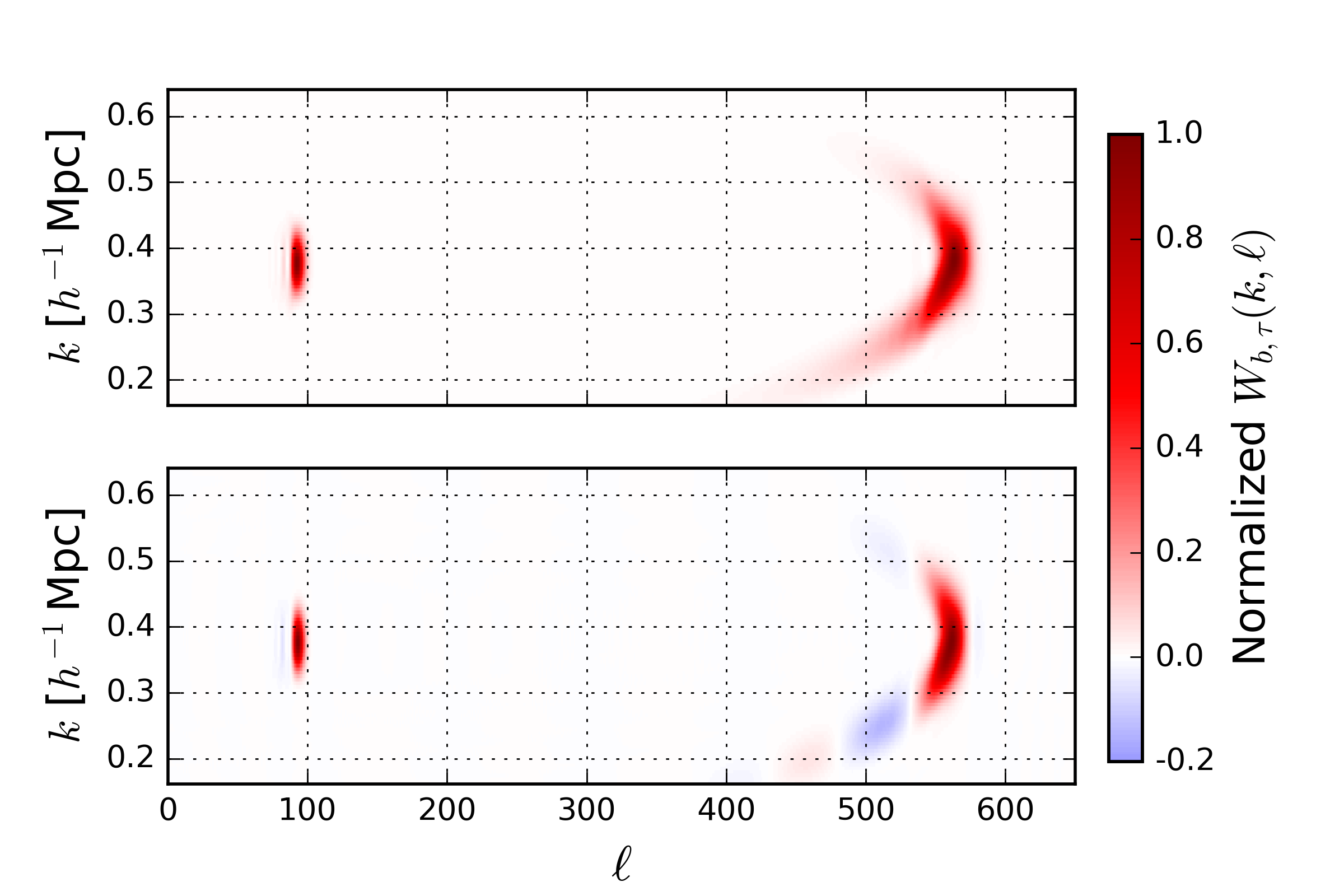}
\caption{Sample window functions for delay-spectrum estimates of the power spectrum explored in this paper. The top plot shows window functions for a traditional equivalent-baseline delay-spectrum estimate, where a single baseline's delay spectrum is squared. The bottom plot shows the window functions resulting from nearly equivalent baselines. In each plot, the window function on the right corresponds to the window function for baselines of length $b \sim 180\,\textrm{m}$ in the East West direction, while the window function on the left is for $b\sim 30\,\textrm{m}$. The general extent of the window functions for the two-baseline estimator are seen to be roughly the same as for the single-baseline estimator, suggesting that the former is just as good as the latter in keeping the EoR window free of foregrounds.}
\label{fig:multiblwindows}
\end{figure}

Figure \ref{fig:multiblwindows} shows four example window functions. The two window functions in the top plot are for equivalent baseline power-spectrum estimators, i.e., where one squares the delay spectrum from a single baseline and normalizes the result. The window function localized at low $\ell$ is for the baseline \{2,0\} of around $30\,\textrm{m}$ in length, while the window function localized at high $\ell$ is for \{12,0\}, roughly $180\,\textrm{m}$ in length. The window functions for near-equivalent baseline estimates are shown in the bottom plot, for \{2,0:2,1\} and \{12,0:12,1\} at low and high $\bar{b}$, respectively. In all cases, the window functions that we show are for $\tau = 703\,\textrm{ns}$, and are generated using primary beam models for PAPER. For numerical convenience we assumed that the bandpass $\phi_\nu$ (see Eq. \eqref{eq:B}) takes the form of a half-period sine curve that peaks at $0.15\,\textrm{\,GHz}$ and goes to zero $5\,\textrm{\,MHz}$ on either side of the peak.

One sees that all four window functions are fairly localized around specific values of $k$, which to a good approximation are described by Eq. \eqref{eq:kbtau}. This suggests that both equivalent- and near-equivalent-baseline delay-spectrum estimators are good estimators of the power spectrum. For window functions that are centered on higher $\ell$ modes, the window functions for both estimators become elongated in $k$, with stronger tails towards the low $k$ modes. Since the low $k$ modes are where the foregrounds reside, measurements of the high $\ell$ modes therefore mix in more foreground power. This is the phenomenology of the wedge, where at fine angular scales the foreground leakage to higher $k$ modes are more pronounced. Importantly, however, we note that the window functions for the near-equivalent-baseline estimator are no more elongated than for the equivalent-baseline estimator. One may thus conclude that our proposed near-equivalent-baseline estimator does not result in extra foreground contamination of the EoR window.


\begin{figure*}[h!]
\includegraphics[width=0.95\textwidth]{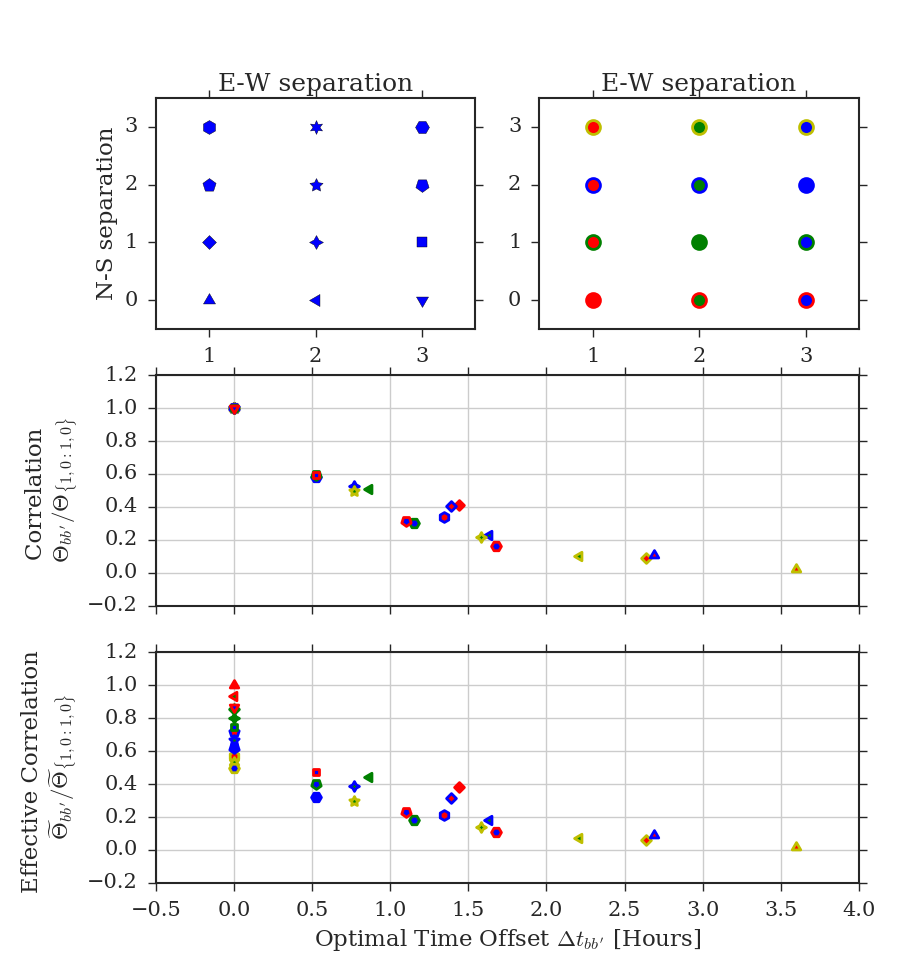}

\caption{Relative sensitivity contributions of selected baseline combinations in PAPER-128. Each labeled point in the middle and bottom panels correspond to a cross-correlated baseline pair \{m,n:p,q\}. The shape of the symbol encodes the first baseline \{m,n\}, as displayed in the top left legend panel. The edge and face colors encode the second baseline \{p,q\}, as displayed in the top right legend panel. The middle
panel shows the peak height ($\Theta$) of each baseline
combination, while the bottom panel multiplies the heights by the
corresponding multiplicities as in Eq. \eqref{eq:sensul}. 
In both the middle and bottom panels, we have chosen to fix the value  of \{1,0:1,0\} to unity.  }
\label{fig:sensplot}
\end{figure*}

\begin{figure}[h]
\includegraphics[width=\linewidth]{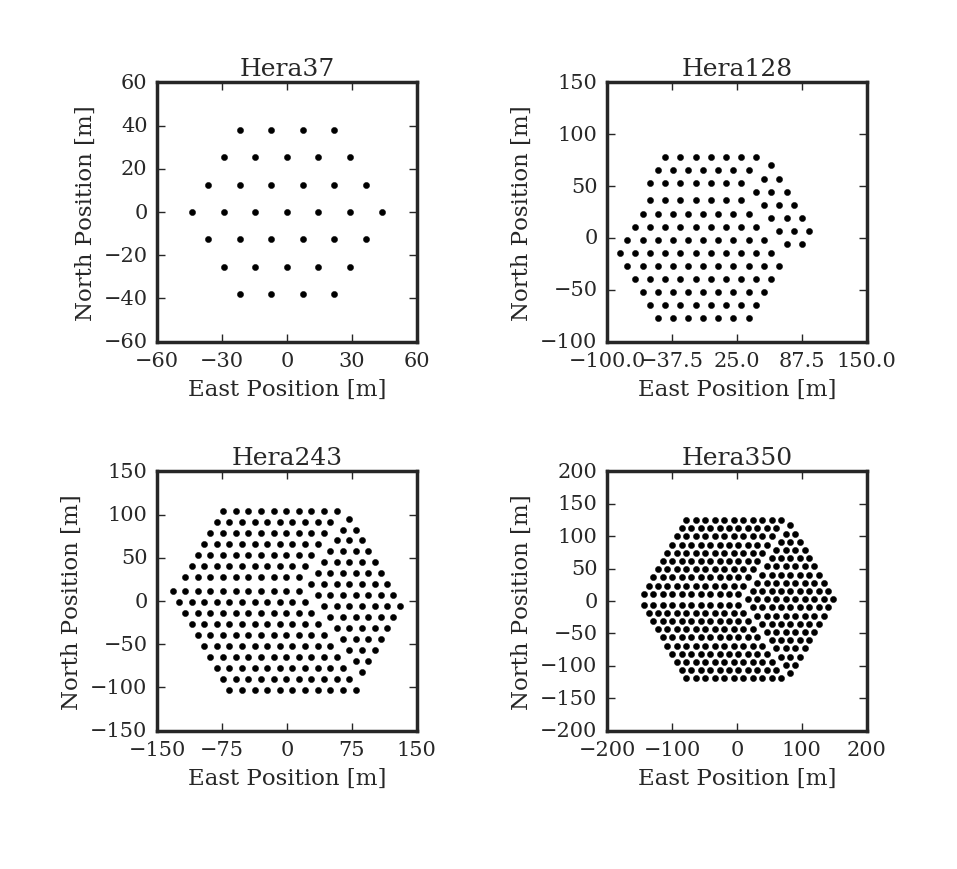}
\label{fig:HeraAntpos}
\caption{Planned Hydrogen Epoch of Reionization Array antenna configurations. HERA-37 is expected to complete and start collecting data in August of 2017, while the other three are planned configurations in the upcoming phases. For HERA-350, only the 320 elements in the core are shown. }
\end{figure}

\subsection{ Sensitivity \label{sec:sensitivity}}
In this section we discuss the sensitivity contributions of a variety of baseline pairs from PAPER-128.
Recall the sensitivity contribution of a particular baseline pair is given by $\Theta$ (Eq. \eqref{eq:Theta}). From now on we use $\Theta_{\boldsymbol{bb'}}$ and $\Delta t_{\boldsymbol{bb'}}$ to denote the peak value and location of $\Theta$, i.e. the maximum of the redundancy weight $|\Theta(\Delta t)|$ and the corresponding offset for a given baseline pair $\{\boldsymbol{b}:\boldsymbol{b'}\}$. 

In the middle panel of Fig. \ref{fig:sensplot}
we show $\Theta_{\boldsymbol{bb'}}$ and $\Delta t_{\boldsymbol{bb'}}$ for a variety of baseline combinations. Baseline pairs that are mirror images of each other 
give approximately the same amount of redundancy ($\Theta_{\boldsymbol{bb'}}$), with the opposite time offset, as expected from symmetry. 
For example, \{1,0:1,1\} is mirror image of \{1,0:1,-1\} and these two pairs of baselines
give the same sensibility contribution. Thus we only show a subset of representative baseline pairs to illustrate the contributions. For a more complete result, see Fig. \ref{fig:pairplot}.  

As is immediately clear from the figure, baseline pairs that have smaller optimal time offsets
tend to have higher correlations. In other words, correlation peaks
with closer to zero time offset are higher. This is expected for two reasons;
one is that the longer the time offset of maximum redundancy, the more the sky has moved with respect
to the primary beams and hence the smaller the overlapping patch of sky surveyed. The other reason is that smaller optimal
time offset corresponds to smaller $w$-term-induced decoherence and hence better redundancy.

To determine the actual relative contribution to sensitivity of these
baseline pairs, we also have to account for the multiplicities of
these baseline classes, i.e., the number of antenna pairs with the
same length and orientation. We would like to estimate an effective weight $\widetilde{\Theta}_{\boldsymbol{bb'}}$ that accounts for both the peak height $\Theta_{\boldsymbol{bb'}}$ and the multiplicities of the baseline classes. From Fig. \ref{fig:sensplot}
we see for example that \{1,0\} has higher multiplicity than \{2,0\},
or \{1,1\}. Assuming that each equivalent baseline delivers
the same $\Theta_{\boldsymbol{bb}}$, the relative contribution to sensitivity can be estimated as follows. 

First we can average the visibilities of the equivalent baselines. Since the core of PAPER-128 was a 16 by 7 antenna configuration, there
are $M\equiv(16-|m|)\times(7-|n|)$ copies of the baseline class $\{m,n\}$. This means
that if we add visibility measurements of all the equivalent baselines,
we get a factor of $\sqrt{M}$ reduction in the noise
level $\sigma_N$ of the visibility. The sensitivity contribution of $\{m,n\}$, cross multiplied with $\{m',n'\}$  thus roughly scales as $\sqrt{(16-|m|)(7-|n|)(16-|m'|)(7-|n'|)}=\sqrt{MM'}$.
For cross-multiplications of nearly equivalent baselines of
types $\{m,n\}$ and $\{m',n'\}$, we get an effective weight: 

\begin{equation}
\widetilde{\Theta}_{\boldsymbol{bb'}} \propto \Theta_{\boldsymbol{bb'}}\sqrt{MM'}.\label{eq:sensul}
\end{equation}

Shown in the bottom panel of Fig. \ref{fig:sensplot} are the peak heights weighted
by the multiplicity factor. Points that have zero optimal time offset are the equivalent baseline pairs and their weighted correlation values simply reflect the multiplicity factor. For clarity of presentation we have ``folded over'' the negative time offsets and combined baseline pairs that are mirror images.  

With $\widetilde{\Theta}_{\boldsymbol{bb'}}$, we can estimate the power spectrum by inverse covariance weighting. We again assume that the power spectrum $P$ varies negligibly over the $k$-space of interest, so that measurements at different $k_{\bar{b}, \tau}$ can be combined to obtain an estimator at some $k_\tau$:
\begin{equation}
\begin{aligned}
 \hat{P}(k_{\tau}) &= \frac{\sum_{\boldsymbol{bb'}}\hat{P}(k_{\bar{b}, \tau})/\sigma_P^2(\boldsymbol{bb'})}{\sum_{\boldsymbol{bb'}}1/\sigma_P^2(\boldsymbol{bb'})}\\
 &= \frac{\sum_{\boldsymbol{bb'}}\hat{P}(k_{\bar{b}, \tau})\widetilde{\Theta}_{\boldsymbol{bb'}}^2}{\sum_{\boldsymbol{bb'}}\widetilde{\Theta}_{\boldsymbol{bb'}}^2},
 \end{aligned}
\end{equation}
where the sum is over classes of baseline pairs, and the power spectrum noise covariance, $\sigma_P^2$, is proportional to the effective weight, $\widetilde{\Theta}$.

We define the estimator sensitivity to be the square root of the inverse of the total power spectrum noise variance $\Sigma_P^2$:
\begin{equation}
\rho \propto 1/\Sigma_P \propto \rho_0\sqrt{\sum_{\boldsymbol{bb'}}^N\widetilde{\Theta}^2_{\boldsymbol{bb'}}},
\end{equation}
where, if $\sigma_S^2$ and $\sigma_N^2$ are the characteristic signal and noise levels of a single-baseline visibility, $\rho_0\equiv\sigma_S^2/\sigma_N^2$ is the signal to noise ratio.

The scaling in Eq. \eqref{eq:sensul} was an approximate one for simplicity of motivation. As we derive in Appendix \ref{sec:appB}, this weight should be corrected by a factor proportional to $\rho_0$:

\begin{equation}
\label{eq:tildereal}
\widetilde{\Theta}_{\boldsymbol{bb'}} =\frac{\Theta_{\boldsymbol{bb'}}\sqrt{MM'}}{\sqrt{1 + \rho_0 \left(M+M' \right)}}.
\end{equation}

For a given $\rho_0$, Eq. \eqref{eq:tildereal} quantifies the relative sensitivity contribution of a baseline pair $\{\boldsymbol{b}:\boldsymbol{b'}\}$. Assuming a reionization signal of $\sim 30\,\text{mK}$, observation centered at 0.15\,GHz ($z=8.5$), and 120 days of integration with PAPER antennas, we have roughly
(see Eq.(20) in \citealt{first-paper})
\begin{equation}
\rho_0 \sim 0.001\left[\frac{\bar{b}}{40\,\text{m}}\right] \left[\frac{0.1h\,\text{Mpc}^{-1}}{k}\right]^3, 
\end{equation}
where $\bar{b}$ is the average baseline length between the pair.  

\subsection{Array Configuration Comparisons \label{sec:arrconf}}
We run our algorithm over all possible baseline-pairs of  PAPER-128, HERA-37, HERA-128, HERA-243 and HERA-350. The HERA antenna configurations are shown in Fig. \ref{fig:HeraAntpos}. The  
hexagonal design is the densest pattern of antenna-packing. The larger arrays are fractured with a ``gap'' dividing the antennas into three different groups. The gaps are designed so as to improve $uv$ coverage and ease calibration without compromising sensitivity, but also produces many more nearly equivalent baselines than a pure hexagonal layout. The motivations behind the designs are explained in \cite{JoshAntPos}.  Compared to PAPER-128, the hexagonal pattern of HERA lack short baselines oriented close to each other, and thus we expect to see only longer nearly equivalent baselines with high correlations. The lower multiplicities per class of baselines is compensated by the larger number of classes of baseline-pairs, especially given the gap in the larger versions. 

\begin{figure}[H]
\includegraphics[width=\linewidth]{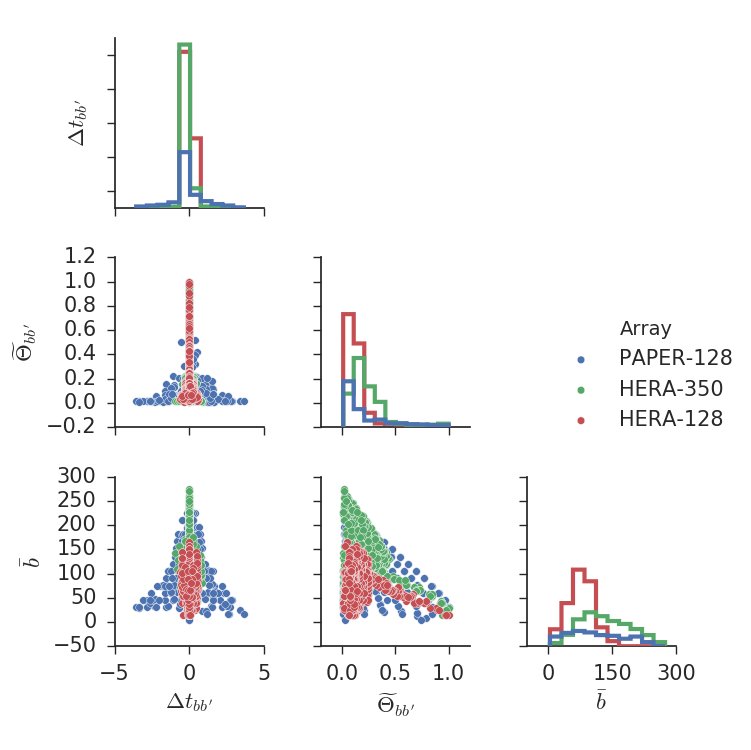}

\caption{Pairplots of the top contributing baseline pairs in three arrays. Plotted properties are optimal time offset $\Delta t_{\boldsymbol{bb'}}$ (in hours),  effective weight $\widetilde{\Theta}_{\boldsymbol{bb'}}$, and baseline length $\bar{b}$ (in meters). On the diagonals we show histograms of given quantities for the different arrays, with the vertical axes linear in scale. On the off-diagonals we show the the scatter plots of the three quantities against each other. Only those points with $\widetilde{\Theta}_{\boldsymbol{bb'}}>0.01$ are shown (the weight for the top class of equivalent pair is normalized to 1). Scatter points of the three arrays overlap, in the order indicated by the legend. }
\label{fig:pairplot}
\end{figure}

\begin{figure}[H]
\includegraphics[width=\linewidth]{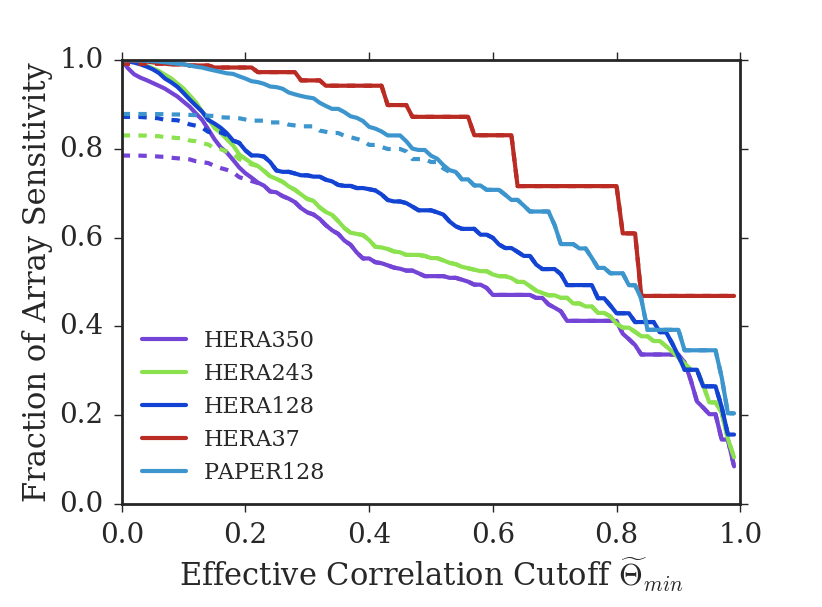}
\caption{Fractional sensitivity of redundant arrays as a function of the minimum effective weight. Dashed lines represent when only equivalent baseline pairs are used, while solid lines indicate use of both the equivalent and nearly equivalent baselines are used. The vertical axis is normalized independently for each array to the total sensitivity when all baseline pairs, equivalent and near-equivalent, are used.}
\label{fig:osens}
\end{figure}

In Fig. \ref{fig:pairplot} we present pair-distributions of three different properties for baselines that contribute well to the sensitivity ($\widetilde{\Theta}_{\boldsymbol{bb'}}>0.01$, where again the weight for the top class of equivalent pair is normalized to 1). The properties shown are effective weight $\widetilde{\Theta}_{\boldsymbol{bb'}}$, optimal time offset $\Delta t_{\boldsymbol{bb'}}$, and average baseline length $\bar{b}$. We show the histograms on the diagonals and scatter plots on the off-diagonals. We only show three of the mentioned arrays for visual clarity. The other two results are qualitatively similar. We point out some key features:
\begin{itemize}

\item $\Delta t_{\boldsymbol{bb'}}$ vs. $\widetilde{\Theta}_{\boldsymbol{bb'}}$:
This relation  is familiar from Fig. \ref{fig:sensplot}. The points at $\Delta t=0$ are the equivalent baselines. HERA arrays have fewer data points with high $\Delta t_{\boldsymbol{bb'}}$, due to their comparatively narrow beam. All near-equivalent pairs for HERA arrays appear at $\widetilde{\Theta}_{\boldsymbol{bb'}}\lesssim 0.4$.

\item $\Delta t_{\boldsymbol{bb'}}$ vs. $\bar{b}$:
PAPER-128 shows the trend that longer baselines correspond to lower $\Delta t_{\boldsymbol{bb'}}$. HERA arrays do not exhibit this trend here because of a selection effect. Shorter HERA baselines require much longer $\Delta t$ to overlap, partly because of the hexagonal structure requires $60\degree$ of rotation to overlap baselines, partly because of the smaller primary beam. Thus most short HERA baseline pairs are not shown in this figure, as they do not meet the selection criterion $\widetilde{\Theta}_{\boldsymbol{bb'}}>0.01$.

\item $\widetilde{\Theta}_{\boldsymbol{bb'}}$ vs. $\bar{b}$:
In this plot the HERA arrays each show two superimposed structures.  The equivalent baselines, all having the same individual weight $\Theta_{\boldsymbol{bb'}}$, appear as elongated structures towards high $\widetilde{\Theta}_{\boldsymbol{bb'}}$. The narrowness of these structures is an indication of the approximately linear and one-to-one correspondence between baseline length and multiplicity. The wide structures spanning the whole range of $\bar{b}$ are the near-equivalent baseline pairs. We see that for HERA arrays, all near-equivalent pairs appear at $\widetilde{\Theta}_{\boldsymbol{bb'}}\lesssim 0.4$. For PAPER-128, the existence of short near-equivalent pairs with high multiplicity partially fills the high $\widetilde{\Theta}_{\boldsymbol{bb'}}$ region. 
Note the general trend that longer baselines tend to have lower $\widetilde{\Theta}_{\boldsymbol{bb'}}$. This is due to the lower multiplicity of longer baselines. The trend is particularly obvious in the linear structure for the HERA arrays, which are the equivalent baselines. All having unity $\Theta_{\boldsymbol{bb'}}$ by definition, the linearly decreasing trend of $\widetilde{\Theta}_{\boldsymbol{bb'}}$ is a direct measure of the baseline multiplicity structures of the HERA array configurations.

\item The top sensitivity-contributing near-equivalent pair in each array is normalized to have the same $\Theta_{\boldsymbol{bb'}}$ (not shown), but those of HERA arrays have much lower $\widetilde{\Theta}_{\boldsymbol{bb'}}$ than the top pair of PAPER-128. This is because they are longer baselines with lower multiplicity. In the end, these baseline classes still lead to high contributions to total sensitivity (Fig. \ref{fig:osens}) because there are many more such baseline pairs for HERA.  

\end{itemize}

As Fig. \ref{fig:pairplot} suggests, not all baseline pairs contribute significantly to sensitivity. Due to the large number of baseline pairs, it may not be computationally feasible to include all pairs. Having quantified the sensitivity from a given pair of baselines, we study the cumulative sensitivity of the array depending on which baseline pairs we include. By construction, we prefer the pairs with larger $\widetilde{\Theta}_{\boldsymbol{bb'}}$. In Fig. \ref{fig:osens}, we plot fractional sensitivity $\rho$ as a function of $\widetilde{\Theta}_{min}$, the minimum cutoff for $\widetilde{\Theta}_{\boldsymbol{bb'}}$. In other words $\rho(\widetilde{\Theta}_{min})$ is the sensitivity when all baseline pairs with $\widetilde{\Theta}_{\boldsymbol{bb'}}>\widetilde{\Theta}_{min}$ are included. 

Note that here we normalized sensitivity for each array as the fraction of the total sensitivity of an array, i.e. when all baseline pairs are used. The plot therefore does not compare absolute sensitivity across different arrays. We see as expected that in all cases, using the nearly equivalent baselines leads to increasingly significant improvements as $\widetilde{\Theta}_{min}$ is lowered, or in other words, when more baseline pairs are used. We point out that PAPER results until now \citep{Ali2015, paper32} only used the top contributing baseline pairs, essentially only exploiting the sensitivity of the $\rho(\widetilde{\Theta}_{min}\sim 1)$ scenario. 

The dashed lines represent the values when only the equivalent baseline-pairs are used. The small and unfractured HERA-37, with no gap (like in HERA-350) or short nearly equivalent baselines (like in PAPER 128), will not benefit much from the nearly equivalent baselines. Near-equivalent baselines make an increasingly significant contribution to the total sensitivity as the number of array elements are increased. PAPER-128 is designed with highly redundant nearly equivalent baselines, and thus  these baselines start contributing at higher $\widetilde{\Theta}_{min}$. As seen in the scatter plot of $\widetilde{\Theta}_{\boldsymbol{bb'}}$ vs. $\bar{b}$ in Fig. \ref{fig:pairplot}, these are also the shorter baselines, which tend to contribute sensitivity to lower $k$-modes. The fractured HERA configurations will benefit even more from nearly equivalent baselines at low $\widetilde{\Theta}_{min}$ due to the presence of more classes of such pairs. However, compared to PAPER, HERA's boost in sensitivity is mostly attributable to the longer baselines---and thus high $k$-modes (see Fig. \ref{fig:pairplot}). The stepwise pattern seen in almost all cases are characteristic of a regular grid; as we step to lower $\widetilde{\Theta}_{\boldsymbol{bb'}}$ large groups of baseline pair classes get included in ``batch''. Thus the ``unfractured'' HERA-37, with more regularity in its antenna configurations than the larger counterparts, exhibit more step patterns in fractional sensitivity. 


\section{Conclusion}
Upcoming arrays aiming at measuring the 21\,cm power spectrum, including HERA and SKA-low, feature increasing number of antennas on regular grids to maximize coherent information. Having efficient pipelines to extract full sensitivity of these arrays is thus a pressing necessity.
 
Current visibility-based power spectrum pipelines have the advantage of easy-to-manage systematics, but have yet to unlock the full sensitivity of the arrays as they do not incorporate partially redundant baselines.  We present a visibility and delay-transform based method to overcome this limitation by extracting the sensitivity contained in such partial redundancy. Applicable to both tracking and drift scan measurements, our method relies on cross-multiplications of visibilities that are offset in time. Given an antenna array configuration, our method identifies the best baseline pairs to cross-multiply and predicts the optimal time offset $\Delta t_{\boldsymbol{bb'}}$, weight $\Theta_{\boldsymbol{bb'}}$, and the necessary rephasing factor $\psi_\nu$. With the predicted results one can incorporate partial redundancy into existing delay-transform based power-spectrum pipelines. 

One benefit of delay-transform based power-spectrum estimation is the ability to isolate foregrounds in Fourier space. We show that our proposed estimator does not lead to extra foreground leakage into the EoR window. We then apply our estimator to a variety array configurations. We show that for arrays with increasing number of antennas, partially redundant baselines make up a more significant portion of the array's total sensitivity. The proposed power-spectrum estimator enables one to unlock the full sensitivity of a redundant radio interferometer while retaining the benefits of delay-transform techniques for isolating foregrounds and avoiding systematics.

\acknowledgments

The authors thank Zaki Ali, Carina Cheng, Dave DeBoer, and Miguel Morales for helpful discussions. This research was completed as part of the University of California Cosmic Dawn Initiative, with support from the University of California Office of the President Multicampus Research Programs and Initiative through award MR-15-328388. Additional support was provided by NSF AST grants 1129258 and 1636646. AL acknowledges support for this work by NASA through Hubble Fellowship grant \#HST-HF2-51363.001-A awarded by the Space Telescope Science Institute, which is operated by the Association of Universities for Research in Astronomy, Inc., for NASA, under contract NAS5-26555.

\appendix
\section{\\Track-crossing and $w$-term \label{sec:appA}}
\label{sec:appA}
In this section we link our results from Section \ref{sec:method} to the traditional views of rotation synthesis and in particular $uv$ tracks. In describing the relative motion of the baseline relative to the sky, we have in Section \ref{sec:method} rotated the direction coordinate $\hat{s}$ with $\boldsymbol{\Gamma}$. 
The typical convention of rotation synthesis has the rotation operator acting on $\boldsymbol{b'}$ instead of $\hat{\boldsymbol{s}}$, in which case Eq. \eqref{eq:Theta} takes the equivalent form
\begin{equation}
\Theta \equiv\int\frac{d\Omega d\nu}{X^{2}Y}B^{*}(\hat{\boldsymbol{s}},\nu)B(\boldsymbol{\Gamma}\hat{\boldsymbol{s}},\nu) e^{-i2\pi\frac{\nu}{c}\hat{\boldsymbol{s}}\cdot\left(\boldsymbol{b}-\boldsymbol{\Gamma}^{-1}\boldsymbol{b'}\right)}, 
\label{eq:drift}
\end{equation}
where $\boldsymbol{\Gamma}^{-1}$ is the inverse of $\boldsymbol{\Gamma}$. 
And for tracking elements we have analogously:
\begin{equation}
\Theta \equiv\int\frac{d\Omega d\nu}{X^{2}Y}B^{*}(\hat{\boldsymbol{s}},\nu)B(\hat{\boldsymbol{s}},\nu) e^{-i2\pi\frac{\nu}{c}\hat{\boldsymbol{s}}\cdot\left(\boldsymbol{b}-\boldsymbol{\Gamma}^{-1}\boldsymbol{b'}\right)}, 
\label{eq:tracking}
\end{equation}
Track-crossing corresponds to vanishing of the exponent
\[
 \boldsymbol{b}-\boldsymbol{\Gamma}^{-1}\boldsymbol{b'} = 0. \label{eq:trackcross}
 \]
If exponent vanishes, $\Theta_{\nu}(\Delta t)$ is real, and one would not observe any non-zero phase at its peak value. However, with the inclusion of the $w$-term, we see that track crossings in $uv$-plane do not actually imply crossing in the $uvw$ space, and therefore the exponent or, equivalently, Eq. \eqref{eq:trackcross} does not vanish and we observe non-zero, frequency dependent peak phases even in the case of tracking arrays, as shown in Fig. \ref{fig:phi_nu}. 

Furthermore, in the case of drift-scan arrays, even a crossing in $uvw$ space does not necessarily correspond to the maximum of correlation. From Eq. \eqref{eq:drift}, we see the track-crossing condition Eq. \eqref{eq:trackcross} maximizes $\Theta$ if and only if no other term in the integral depends on $\boldsymbol{\Gamma}$. We see that this is true for the tracking case (Eq. \eqref{eq:tracking}), but not the drift-scan case (Eq. \eqref{eq:drift}). Only in the special case of a tracking measurement, where the $w$-term happens to be negligible, do $uv$-track crossings correspond exactly to maxima of correlations. In this case, at the crossing points the exponent in Eq. \eqref{eq:trackcross} vanishes and $\psi_\nu$ is zero for all $\nu$. In general, second-order effects due to the presence of $w$ are observed for the tracking measurements as in Fig. \ref{fig:phi_nu}.

\section{\label{sec:appB}\\Derivation of Noise Covariance \label{sec:appB}}
\label{sec:appB}
In this appendix we provide a derivation of the effective weight $\widetilde{\Theta}$ quoted in Section \ref{sec:sensitivity}. In doing so, we assume that the power spectrum measurements from distinct baseline classes are combined by inverse variance weighting. To begin, we separate the visibility and power spectrum into signal and noise contributions:
\begin{equation}
\begin{aligned}
V &= V_S+V_N,\\
P &= P_S+P_N.
\end{aligned}
\end{equation}
We write the noise variance of power spectrum and visibility as:
\begin{equation}
\begin{aligned}
\sigma_V^2 &= \langle |V_N|^2 \rangle,\\
\sigma_P^2 &= \langle P_N^2 \rangle.
\end{aligned}
\end{equation}
One may notice that we use a single covariance for the complex visibility. It is straightforward to show that the same result holds if a separate real and imaginary components are used, as long as they are independent of each other. In fact, for simplicity and without loss of generality we shall treat the visibility as a real quantity in the rest of this derivation. 
Note that though we can assume $\langle V_N^{m}\rangle=0$, where $m$ is odd. The same is not true for $P_N$. 

The variance of $P$ constructed with visibilities $V_1$ and $V_2$ from two baseline classes can be estimated:
\begin{equation}
\begin{aligned}
\sigma_P^2 &= \langle P^2\rangle -\langle P \rangle^2,\\
&\propto \langle \frac{(V_{1S}+V_{1N})^2 (V_{2S}+V_{2N})^2}{\Theta^2} \rangle - \langle \frac{(V_{1S}+V_{1N}) (V_{2S}+V_{2N})}{\Theta} \rangle ^2\\
&= \frac{1}{\Theta_{12}^2} \left( V_{1S}^2\sigma_{V2}^2+V_{2S}^2\sigma_{V1}^2+\langle V_{1N}^2 V_{2N}^2\rangle\right)\\
&= \frac{1}{\Theta_{12}^2} \left[ V_{S}^2(\sigma_{V2}^2+\sigma_{V1}^2) + \sigma_{V1}^2 \sigma_{V2}^2\right], 
\end{aligned}
\end{equation}
where in the second last line we have substituted in the visibility noise variance, and have assumed independent noise between baselines. In the final line we used Wick's theorem and the fact that the signal from the two visibilities are equal. 

Recall from the discussion on multiplicities that we can write
\begin{equation}
\sigma_V^2=\frac{\sigma_0^2}{M},
\end{equation}
where $\sigma_0$ is the single-baseline noise level. Letting $\rho_0=V_S^2/\sigma_0^2$ be the signal to noise ratio for a single baseline, we can write

\begin{equation}
\begin{aligned}
\sigma_P^2 & \propto  \frac{\sigma_0^4}{\Theta_{12}^2} \left[ \rho_0 \left(\frac{1}{M_1}+\frac{1}{M_2} \right) + \frac{1}{M_1 M_2}\right]\\
&\propto \frac{1}{\widetilde{\Theta}_{12}^2}.
\end{aligned}
\end{equation}
Thus we have defined a slightly modified version of the effective weight in Eq.  \eqref{eq:tildereal}:
\begin{equation}
\widetilde{\Theta}_{12}=\frac{\Theta_{12}\sqrt{M_1M_{2}}}{\sqrt{1 + \rho_0 \left(M_1+M_{2} \right)}}.
\end{equation}

\bibliographystyle{apj}
\bibliography{draft_working}

\begin{thebibliography}{57}
\expandafter\ifx\csname natexlab\endcsname\relax\def\natexlab#1{#1}\fi

\bibitem[{Ali {et~al.}(2015)Ali, Parsons, Zheng, Pober, Liu, Aguirre, Bradley,
  Bernardi, Carilli, Cheng, DeBoer, Dexter, Grobbelaar, Horrell, Jacobs, Klima,
  MacMahon, Maree, Moore, Razavi, Stefan, Walbrugh, \& Walker}]{Ali2015}
Ali, Z.~S., {et~al.} 2015, \apj, 809, 61

\bibitem[{{Beardsley} {et~al.}(2016){Beardsley}, {Hazelton}, {Sullivan},
  {Carroll}, {Barry}, {Rahimi}, {Pindor}, {Trott}, {Line}, {Jacobs}, {Morales},
  {Pober}, {Bernardi}, {Bowman}, {Busch}, {Briggs}, {Cappallo}, {Corey}, {de
  Oliveira-Costa}, {Dillon}, {Emrich}, {Ewall-Wice}, {Feng}, {Gaensler},
  {Goeke}, {Greenhill}, {Hewitt}, {Hurley-Walker}, {Johnston-Hollitt},
  {Kaplan}, {Kasper}, {Kim}, {Kratzenberg}, {Lenc}, {Loeb}, {Lonsdale},
  {Lynch}, {McKinley}, {McWhirter}, {Mitchell}, {Morgan}, {Neben},
  {Thyagarajan}, {Oberoi}, {Offringa}, {Ord}, {Paul}, {Prabu}, {Procopio},
  {Riding}, {Rogers}, {Roshi}, {Udaya Shankar}, {Sethi}, {Srivani},
  {Subrahmanyan}, {Tegmark}, {Tingay}, {Waterson}, {Wayth}, {Webster},
  {Whitney}, {Williams}, {Williams}, {Wu}, \& {Wyithe}}]{MWAresult1}
{Beardsley}, A.~P., {et~al.} 2016, \apj, 833, 102

\bibitem[{Bowman {et~al.}(2013)Bowman, Cairns, Kaplan, Murphy, Oberoi,
  Staveley-Smith, Arcus, Barnes, Bernardi, Briggs, \& et~al.}]{Bowman2013}
Bowman, J.~D., {et~al.} 2013, \pasa, 30

\bibitem[{Bull {et~al.}(2015)Bull, Ferreira, Patel, \& Santos}]{Bull2015}
Bull, P., Ferreira, P.~G., Patel, P., \& Santos, M.~G. 2015, \apj, 803, 21

\bibitem[{{Chapman} {et~al.}(2016){Chapman}, {Zaroubi}, {Abdalla}, {Dulwich},
  {Jeli{\'c}}, \& {Mort}}]{chapman_et_al2016}
{Chapman}, E., {Zaroubi}, S., {Abdalla}, F.~B., {Dulwich}, F., {Jeli{\'c}}, V.,
  \& {Mort}, B. 2016, \mnras, 458, 2928

\bibitem[{Chen(2015)}]{DEw21cm}
Chen, X. 2015, International Journal of Modern Physics A, 30, 1545011

\bibitem[{{Choudhuri} {et~al.}(2016{\natexlab{a}}){Choudhuri}, {Bharadwaj},
  {Chatterjee}, {Ali}, {Roy}, \& {Ghosh}}]{Choudhury2016Taperingb}
{Choudhuri}, S., {Bharadwaj}, S., {Chatterjee}, S., {Ali}, S.~S., {Roy}, N., \&
  {Ghosh}, A. 2016{\natexlab{a}}, \mnras, 463, 4093

\bibitem[{{Choudhuri} {et~al.}(2014){Choudhuri}, {Bharadwaj}, {Ghosh}, \&
  {Ali}}]{Choudhury2014Estimators}
{Choudhuri}, S., {Bharadwaj}, S., {Ghosh}, A., \& {Ali}, S.~S. 2014, \mnras,
  445, 4351

\bibitem[{{Choudhuri} {et~al.}(2016{\natexlab{b}}){Choudhuri}, {Bharadwaj},
  {Roy}, {Ghosh}, \& {Ali}}]{Choudhury2016Taperinga}
{Choudhuri}, S., {Bharadwaj}, S., {Roy}, N., {Ghosh}, A., \& {Ali}, S.~S.
  2016{\natexlab{b}}, \mnras, 459, 151

\bibitem[{{Datta} {et~al.}(2010){Datta}, {Bowman}, \& {Carilli}}]{Datta2010}
{Datta}, A., {Bowman}, J.~D., \& {Carilli}, C.~L. 2010, \apj, 724, 526

\bibitem[{{DeBoer} {et~al.}(2016){DeBoer}, {Parsons}, {Aguirre}, {Alexander},
  {Ali}, {Beardsley}, {Bernardi}, {Bowman}, {Bradley}, {Carilli}, {Cheng}, {de
  Lera Acedo}, {Dillon}, {Ewall-Wice}, {Fadana}, {Fagnoni}, {Fritz},
  {Furlanetto}, {Glendenning}, {Greig}, {Grobbelaar}, {Hazelton}, {Hewitt},
  {Hickish}, {Jacobs}, {Julius}, {Kariseb}, {Kohn}, {Lekalake}, {Liu}, {Loots},
  {MacMahon}, {Malan}, {Malgas}, {Maree}, {Mathison}, {Matsetela}, {Mesinger},
  {Morales}, {Neben}, {Patra}, {Pieterse}, {Pober}, {Razavi-Ghods},
  {Ringuette}, {Robnett}, {Rosie}, {Sell}, {Smith}, {Syce}, {Tegmark},
  {Thyagarajan}, {Williams}, \& {Zheng}}]{HERA}
{DeBoer}, D.~R., {et~al.} 2016, ArXiv e-prints: 1606.07473

\bibitem[{{Dillon} \& {Parsons}(2016)}]{JoshAntPos}
{Dillon}, J.~S., \& {Parsons}, A.~R. 2016, \apj, 826, 181

\bibitem[{{Dillon} {et~al.}(2014){Dillon}, {Liu}, {Williams}, {Hewitt},
  {Tegmark}, {Morgan}, {Levine}, {Morales}, {Tingay}, {Bernardi}, {Bowman},
  {Briggs}, {Cappallo}, {Emrich}, {Mitchell}, {Oberoi}, {Prabu}, {Wayth}, \&
  {Webster}}]{dillon_et_al2014}
{Dillon}, J.~S., {et~al.} 2014, \prd, 89, 023002

\bibitem[{{Dillon} {et~al.}(2015){Dillon}, {Neben}, {Hewitt}, {Tegmark},
  {Barry}, {Beardsley}, {Bowman}, {Briggs}, {Carroll}, {de Oliveira-Costa},
  {Ewall-Wice}, {Feng}, {Greenhill}, {Hazelton}, {Hernquist}, {Hurley-Walker},
  {Jacobs}, {Kim}, {Kittiwisit}, {Lenc}, {Line}, {Loeb}, {McKinley},
  {Mitchell}, {Morales}, {Offringa}, {Paul}, {Pindor}, {Pober}, {Procopio},
  {Riding}, {Sethi}, {Shankar}, {Subrahmanyan}, {Sullivan}, {Thyagarajan},
  {Tingay}, {Trott}, {Wayth}, {Webster}, {Wyithe}, {Bernardi}, {Cappallo},
  {Deshpande}, {Johnston-Hollitt}, {Kaplan}, {Lonsdale}, {McWhirter}, {Morgan},
  {Oberoi}, {Ord}, {Prabu}, {Srivani}, {Williams}, \& {Williams}}]{MWAresult0}
---. 2015, \prd, 91, 123011

\bibitem[{{Ewall-Wice} {et~al.}(2016){Ewall-Wice}, {Bradley}, {DeBoer},
  {Hewitt}, {Parsons}, {Aguirre}, {Ali}, {Bowman}, {Cheng}, {Neben}, {Patra},
  {Thyagarajan}, {Venter}, {de Lera Acedo}, {Dillon}, {Doolittle}, {Egan},
  {Hendrick}, {Klima}, {Kohn}, {Schaffner}, {Shelton}, {Saliwanchik},
  {Tegmark}, {Taylor}, {Taylor}, \& {Wirt}}]{HERADISH2}
{Ewall-Wice}, A., {et~al.} 2016, ArXiv e-prints: 1602.06277

\bibitem[{{Fan} {et~al.}(2006){Fan}, {Carilli}, \& {Keating}}]{Fan2006}
{Fan}, X., {Carilli}, C.~L., \& {Keating}, B. 2006, \araa, 44, 415

\bibitem[{Furlanetto {et~al.}(2006)Furlanetto, Oh, \&
  Briggs}]{Furlanetto2006181}
Furlanetto, S.~R., Oh, S.~P., \& Briggs, F.~H. 2006, Physics Reports, 433, 181

\bibitem[{{George} {et~al.}(2015){George}, {Reichardt}, {Aird}, {Benson},
  {Bleem}, {Carlstrom}, {Chang}, {Cho}, {Crawford}, {Crites}, {de Haan},
  {Dobbs}, {Dudley}, {Halverson}, {Harrington}, {Holder}, {Holzapfel}, {Hou},
  {Hrubes}, {Keisler}, {Knox}, {Lee}, {Leitch}, {Lueker}, {Luong-Van},
  {McMahon}, {Mehl}, {Meyer}, {Millea}, {Mocanu}, {Mohr}, {Montroy}, {Padin},
  {Plagge}, {Pryke}, {Ruhl}, {Schaffer}, {Shaw}, {Shirokoff}, {Spieler},
  {Staniszewski}, {Stark}, {Story}, {van Engelen}, {Vanderlinde}, {Vieira},
  {Williamson}, \& {Zahn}}]{GeorgeKSZ}
{George}, E.~M., {et~al.} 2015, \apj, 799, 177

\bibitem[{{Hazelton} {et~al.}(2013){Hazelton}, {Morales}, \&
  {Sullivan}}]{Hazelton2013}
{Hazelton}, B.~J., {Morales}, M.~F., \& {Sullivan}, I.~S. 2013, \apj, 770, 156

\bibitem[{{Jensen} {et~al.}(2016){Jensen}, {Majumdar}, {Mellema}, {Lidz},
  {Iliev}, \& {Dixon}}]{jensen_et_al2016}
{Jensen}, H., {Majumdar}, S., {Mellema}, G., {Lidz}, A., {Iliev}, I.~T., \&
  {Dixon}, K.~L. 2016, \mnras, 456, 66

\bibitem[{{Kohn} {et~al.}(2016){Kohn}, {Aguirre}, {Nunhokee}, {Bernardi},
  {Pober}, {Ali}, {Bradley}, {Carilli}, {DeBoer}, {Gugliucci}, {Jacobs},
  {Klima}, {MacMahon}, {Manley}, {Moore}, {Parsons}, {Stefan}, \&
  {Walbrugh}}]{kohn_et_al2016}
{Kohn}, S.~A., {et~al.} 2016, \apj, 823, 88

\bibitem[{{Koopmans} {et~al.}(2015){Koopmans}, {Pritchard}, {Mellema},
  {Aguirre}, {Ahn}, {Barkana}, {van Bemmel}, {Bernardi}, {Bonaldi}, {Briggs},
  {de Bruyn}, {Chang}, {Chapman}, {Chen}, {Ciardi}, {Dayal}, {Ferrara},
  {Fialkov}, {Fiore}, {Ichiki}, {Illiev}, {Inoue}, {Jelic}, {Jones}, {Lazio},
  {Maio}, {Majumdar}, {Mack}, {Mesinger}, {Morales}, {Parsons}, {Pen},
  {Santos}, {Schneider}, {Semelin}, {de Souza}, {Subrahmanyan}, {Takeuchi},
  {Vedantham}, {Wagg}, {Webster}, {Wyithe}, {Datta}, \& {Trott}}]{skalow2015}
{Koopmans}, L., {et~al.} 2015, Advancing Astrophysics with the Square Kilometre
  Array (AASKA14), 1

\bibitem[{Liu \& Parsons(2016)}]{Liu2016b}
Liu, A., \& Parsons, A.~R. 2016, MNRAS, 457, 1864

\bibitem[{Liu {et~al.}(2014{\natexlab{a}})Liu, Parsons, \& Trott}]{wedge1}
Liu, A., Parsons, A.~R., \& Trott, C.~M. 2014{\natexlab{a}}, \prd, 90, 023018

\bibitem[{Liu {et~al.}(2014{\natexlab{b}})Liu, Parsons, \& Trott}]{wedge2}
---. 2014{\natexlab{b}}, \prd, 90, 023019

\bibitem[{Liu {et~al.}(2016)Liu, Pritchard, Allison, Parsons, Seljak, \&
  Sherwin}]{LiuOpticalDepth}
Liu, A., Pritchard, J.~R., Allison, R., Parsons, A.~R., Seljak, U., \& Sherwin,
  B.~D. 2016, \prd, 93, 043013

\bibitem[{{Liu} {et~al.}(2016){Liu}, {Zhang}, \& {Parsons}}]{sphere}
{Liu}, A., {Zhang}, Y., \& {Parsons}, A.~R. 2016, \apj, 833, 242

\bibitem[{Mao {et~al.}(2008)Mao, Tegmark, McQuinn, Zaldarriaga, \&
  Zahn}]{Mao2008}
Mao, Y., Tegmark, M., McQuinn, M., Zaldarriaga, M., \& Zahn, O. 2008, \prd, 78,
  023529

\bibitem[{{McQuinn} {et~al.}(2007){McQuinn}, {Hernquist}, {Zaldarriaga}, \&
  {Dutta}}]{mcquinnLyA}
{McQuinn}, M., {Hernquist}, L., {Zaldarriaga}, M., \& {Dutta}, S. 2007, \mnras,
  381, 75

\bibitem[{{Morales} {et~al.}(2012){Morales}, {Hazelton}, {Sullivan}, \&
  {Beardsley}}]{Morales2012}
{Morales}, M.~F., {Hazelton}, B., {Sullivan}, I., \& {Beardsley}, A. 2012,
  \apj, 752, 137

\bibitem[{{Neben} {et~al.}(2016){Neben}, {Bradley}, {Hewitt}, {DeBoer},
  {Parsons}, {Aguirre}, {Ali}, {Cheng}, {Ewall-Wice}, {Patra}, {Thyagarajan},
  {Bowman}, {Dickenson}, {Dillon}, {Doolittle}, {Egan}, {Hedrick}, {Jacobs},
  {Kohn}, {Klima}, {Moodley}, {Saliwanchik}, {Schaffner}, {Shelton}, {Taylor},
  {Taylor}, {Tegmark}, {Wirt}, \& {Zheng}}]{HERABEAM1}
{Neben}, A.~R., {et~al.} 2016, ArXiv e-prints: 1602.03887

\bibitem[{Oyama {et~al.}(2013)Oyama, Shimizu, \& Kohri}]{Oyama20131186}
Oyama, Y., Shimizu, A., \& Kohri, K. 2013, Physics Letters B, 718, 1186

\bibitem[{Parsons {et~al.}(2012{\natexlab{a}})Parsons, Pober, McQuinn, Jacobs,
  \& Aguirre}]{first-paper}
Parsons, A., Pober, J., McQuinn, M., Jacobs, D., \& Aguirre, J.
  2012{\natexlab{a}}, \apj, 753, 81

\bibitem[{{Parsons} \& {Backer}(2009)}]{PB2009}
{Parsons}, A.~R., \& {Backer}, D.~C. 2009, \aj, 138, 219

\bibitem[{Parsons {et~al.}(2012{\natexlab{b}})Parsons, Pober, Aguirre, Carilli,
  Jacobs, \& Moore}]{delay-transform}
Parsons, A.~R., Pober, J.~C., Aguirre, J.~E., Carilli, C.~L., Jacobs, D.~C., \&
  Moore, D.~F. 2012{\natexlab{b}}, \apj, 756, 165

\bibitem[{Parsons {et~al.}(2014)Parsons, Liu, Aguirre, Ali, Bradley, Carilli,
  DeBoer, Dexter, Gugliucci, Jacobs, Klima, MacMahon, Manley, Moore, Pober,
  Stefan, \& Walbrugh}]{paper32}
Parsons, A.~R., {et~al.} 2014, \apj, 788, 106

\bibitem[{{Patil} {et~al.}(2017){Patil}, {Yatawatta}, {Koopmans}, {de Bruyn},
  {Brentjens}, {Zaroubi}, {Asad}, {Hatef}, {Jeli{\'c}}, {Mevius}, {Offringa},
  {Pandey}, {Vedantham}, {Abdalla}, {Brouw}, {Chapman}, {Ciardi}, {Gehlot},
  {Ghosh}, {Harker}, {Iliev}, {Kakiichi}, {Majumdar}, {Mellema}, {Silva},
  {Schaye}, {Vrbanec}, \& {Wijnholds}}]{LOFARresult}
{Patil}, A.~H., {et~al.} 2017, \apj, 838, 65

\bibitem[{{Paul} {et~al.}(2016){Paul}, {Sethi}, {Morales}, {Dwarkanath}, {Udaya
  Shankar}, {Subrahmanyan}, {Barry}, {Beardsley}, {Bowman}, {Briggs},
  {Carroll}, {de Oliveira-Costa}, {Dillon}, {Ewall-Wice}, {Feng}, {Greenhill},
  {Gaensler}, {Hazelton}, {Hewitt}, {Hurley-Walker}, {Jacobs}, {Kim},
  {Kittiwisit}, {Lenc}, {Line}, {Loeb}, {McKinley}, {Mitchell}, {Neben},
  {Offringa}, {Pindor}, {Pober}, {Procopio}, {Riding}, {Sullivan}, {Tegmark},
  {Thyagarajan}, {Tingay}, {Trott}, {Wayth}, {Webster}, {Wyithe}, {Cappallo},
  {Johnston-Hollitt}, {Kaplan}, {Lonsdale}, {McWhirter}, {Morgan}, {Oberoi},
  {Ord}, {Prabu}, {Srivani}, {Williams}, \& {Williams}}]{wterm}
{Paul}, S., {et~al.} 2016, \apj, 833, 213

\bibitem[{{Planck Collaboration} {et~al.}(2016){Planck Collaboration}, {Adam},
  {Aghanim}, {Ashdown}, {Aumont}, {Baccigalupi}, {Ballardini}, {Banday},
  {Barreiro}, {Bartolo}, {Basak}, {Battye}, {Benabed}, {Bernard}, {Bersanelli},
  {Bielewicz}, {Bock}, {Bonaldi}, {Bonavera}, {Bond}, {Borrill}, {Bouchet},
  {Boulanger}, {Bucher}, {Burigana}, {Calabrese}, {Cardoso}, {Carron},
  {Chiang}, {Colombo}, {Combet}, {Comis}, {Couchot}, {Coulais}, {Crill},
  {Curto}, {Cuttaia}, {Davis}, {de Bernardis}, {de Rosa}, {de Zotti},
  {Delabrouille}, {Di Valentino}, {Dickinson}, {Diego}, {Dor{\'e}}, {Douspis},
  {Ducout}, {Dupac}, {Elsner}, {En{\ss}lin}, {Eriksen}, {Falgarone}, {Fantaye},
  {Finelli}, {Forastieri}, {Frailis}, {Fraisse}, {Franceschi}, {Frolov},
  {Galeotta}, {Galli}, {Ganga}, {G{\'e}nova-Santos}, {Gerbino}, {Ghosh},
  {Gonz{\'a}lez-Nuevo}, {G{\'o}rski}, {Gruppuso}, {Gudmundsson}, {Hansen},
  {Helou}, {Henrot-Versill{\'e}}, {Herranz}, {Hivon}, {Huang}, {Ili{\'c}},
  {Jaffe}, {Jones}, {Keih{\"a}nen}, {Keskitalo}, {Kisner}, {Knox},
  {Krachmalnicoff}, {Kunz}, {Kurki-Suonio}, {Lagache}, {L{\"a}hteenm{\"a}ki},
  {Lamarre}, {Langer}, {Lasenby}, {Lattanzi}, {Lawrence}, {Le Jeune},
  {Levrier}, {Lewis}, {Liguori}, {Lilje}, {L{\'o}pez-Caniego}, {Ma},
  {Mac{\'{\i}}as-P{\'e}rez}, {Maggio}, {Mangilli}, {Maris}, {Martin},
  {Mart{\'{\i}}nez-Gonz{\'a}lez}, {Matarrese}, {Mauri}, {McEwen}, {Meinhold},
  {Melchiorri}, {Mennella}, {Migliaccio}, {Miville-Desch{\^e}nes}, {Molinari},
  {Moneti}, {Montier}, {Morgante}, {Moss}, {Naselsky}, {Natoli}, {Oxborrow},
  {Pagano}, {Paoletti}, {Partridge}, {Patanchon}, {Patrizii}, {Perdereau},
  {Perotto}, {Pettorino}, {Piacentini}, {Plaszczynski}, {Polastri}, {Polenta},
  {Puget}, {Rachen}, {Racine}, {Reinecke}, {Remazeilles}, {Renzi}, {Rocha},
  {Rossetti}, {Roudier}, {Rubi{\~n}o-Mart{\'{\i}}n}, {Ruiz-Granados},
  {Salvati}, {Sandri}, {Savelainen}, {Scott}, {Sirri}, {Sunyaev}, {Suur-Uski},
  {Tauber}, {Tenti}, {Toffolatti}, {Tomasi}, {Tristram}, {Trombetti},
  {Valiviita}, {Van Tent}, {Vielva}, {Villa}, {Vittorio}, {Wandelt}, {Wehus},
  {White}, {Zacchei}, \& {Zonca}}]{Planck2016}
{Planck Collaboration} {et~al.} 2016, \aap, 596, A108

\bibitem[{{Pober} {et~al.}(2013){Pober}, {Parsons}, {Aguirre}, {Ali},
  {Bradley}, {Carilli}, {DeBoer}, {Dexter}, {Gugliucci}, {Jacobs}, {Klima},
  {MacMahon}, {Manley}, {Moore}, {Stefan}, \& {Walbrugh}}]{pober_et_al2013b}
{Pober}, J.~C., {et~al.} 2013, \apjl, 768, L36

\bibitem[{{Pober} {et~al.}(2014){Pober}, {Liu}, {Dillon}, {Aguirre}, {Bowman},
  {Bradley}, {Carilli}, {DeBoer}, {Hewitt}, {Jacobs}, {McQuinn}, {Morales},
  {Parsons}, {Tegmark}, \& {Werthimer}}]{Pobersens}
---. 2014, \apj, 782, 66

\bibitem[{{Pober} {et~al.}(2016){Pober}, {Hazelton}, {Beardsley}, {Barry},
  {Martinot}, {Sullivan}, {Morales}, {Bell}, {Bernardi}, {Bhat}, {Bowman},
  {Briggs}, {Cappallo}, {Carroll}, {Corey}, {de Oliveira-Costa}, {Deshpande},
  {Dillon}, {Emrich}, {Ewall-Wice}, {Feng}, {Goeke}, {Greenhill}, {Hewitt},
  {Hindson}, {Hurley-Walker}, {Jacobs}, {Johnston-Hollitt}, {Kaplan}, {Kasper},
  {Kim}, {Kittiwisit}, {Kratzenberg}, {Kudryavtseva}, {Lenc}, {Line}, {Loeb},
  {Lonsdale}, {Lynch}, {McKinley}, {McWhirter}, {Mitchell}, {Morgan}, {Neben},
  {Oberoi}, {Offringa}, {Ord}, {Paul}, {Pindor}, {Prabu}, {Procopio}, {Riding},
  {Rogers}, {Roshi}, {Sethi}, {Udaya Shankar}, {Srivani}, {Subrahmanyan},
  {Tegmark}, {Thyagarajan}, {Tingay}, {Trott}, {Waterson}, {Wayth}, {Webster},
  {Whitney}, {Williams}, {Williams}, \& {Wyithe}}]{pober_et_al2016}
---. 2016, \apj, 819, 8

\bibitem[{Pritchard \& Loeb(2012)}]{PritchardLoeb}
Pritchard, J.~R., \& Loeb, A. 2012, Reports on Progress in Physics, 75, 086901

\bibitem[{{Rhoads} {et~al.}(2012){Rhoads}, {Hibon}, {Malhotra}, {Cooper}, \&
  {Weiner}}]{Rhoads2012}
{Rhoads}, J.~E., {Hibon}, P., {Malhotra}, S., {Cooper}, M., \& {Weiner}, B.
  2012, \apjl, 752, L28

\bibitem[{{Seo} \& {Hirata}(2016)}]{seo_and_hirata2016}
{Seo}, H.-J., \& {Hirata}, C.~M. 2016, \mnras, 456, 3142

\bibitem[{{Smith} \& {Ferraro}(2016)}]{kszpatchy}
{Smith}, K.~M., \& {Ferraro}, S. 2016, ArXiv e-prints: 1607.01769

\bibitem[{{Thomson} {et~al.}(2017){Thomson}, {Moran}, \& {Swenson}}]{TMS}
{Thomson}, A., {Moran}, J., \& {Swenson}, G. 2017, Interferometry and Synthesis
  in Radio Astronomy, 3rd edn. (Springer International Publishing)

\bibitem[{{Thyagarajan} {et~al.}(2013){Thyagarajan}, {Udaya Shankar},
  {Subrahmanyan}, {Arcus}, {Bernardi}, {Bowman}, {Briggs}, {Bunton},
  {Cappallo}, {Corey}, {deSouza}, {Emrich}, {Gaensler}, {Goeke}, {Greenhill},
  {Hazelton}, {Herne}, {Hewitt}, {Johnston-Hollitt}, {Kaplan}, {Kasper},
  {Kincaid}, {Koenig}, {Kratzenberg}, {Lonsdale}, {Lynch}, {McWhirter},
  {Mitchell}, {Morales}, {Morgan}, {Oberoi}, {Ord}, {Pathikulangara},
  {Remillard}, {Rogers}, {Anish Roshi}, {Salah}, {Sault}, {Srivani}, {Stevens},
  {Thiagaraj}, {Tingay}, {Wayth}, {Waterson}, {Webster}, {Whitney}, {Williams},
  {Williams}, \& {Wyithe}}]{Thyagarajan2013}
{Thyagarajan}, N., {et~al.} 2013, \apj, 776, 6

\bibitem[{{Thyagarajan} {et~al.}(2015{\natexlab{a}}){Thyagarajan}, {Jacobs},
  {Bowman}, {Barry}, {Beardsley}, {Bernardi}, {Briggs}, {Cappallo}, {Carroll},
  {Deshpande}, {de Oliveira-Costa}, {Dillon}, {Ewall-Wice}, {Feng},
  {Greenhill}, {Hazelton}, {Hernquist}, {Hewitt}, {Hurley-Walker},
  {Johnston-Hollitt}, {Kaplan}, {Kim}, {Kittiwisit}, {Lenc}, {Line}, {Loeb},
  {Lonsdale}, {McKinley}, {McWhirter}, {Mitchell}, {Morales}, {Morgan},
  {Neben}, {Oberoi}, {Offringa}, {Ord}, {Paul}, {Pindor}, {Pober}, {Prabu},
  {Procopio}, {Riding}, {Udaya Shankar}, {Sethi}, {Srivani}, {Subrahmanyan},
  {Sullivan}, {Tegmark}, {Tingay}, {Trott}, {Wayth}, {Webster}, {Williams},
  {Williams}, \& {Wyithe}}]{Thyagarajan_et_al2015b}
---. 2015{\natexlab{a}}, \apjl, 807, L28

\bibitem[{{Thyagarajan} {et~al.}(2015{\natexlab{b}}){Thyagarajan}, {Jacobs},
  {Bowman}, {Barry}, {Beardsley}, {Bernardi}, {Briggs}, {Cappallo}, {Carroll},
  {Corey}, {de Oliveira-Costa}, {Dillon}, {Emrich}, {Ewall-Wice}, {Feng},
  {Goeke}, {Greenhill}, {Hazelton}, {Hewitt}, {Hurley-Walker},
  {Johnston-Hollitt}, {Kaplan}, {Kasper}, {Kim}, {Kittiwisit}, {Kratzenberg},
  {Lenc}, {Line}, {Loeb}, {Lonsdale}, {Lynch}, {McKinley}, {McWhirter},
  {Mitchell}, {Morales}, {Morgan}, {Neben}, {Oberoi}, {Offringa}, {Ord},
  {Paul}, {Pindor}, {Pober}, {Prabu}, {Procopio}, {Riding}, {Rogers}, {Roshi},
  {Udaya Shankar}, {Sethi}, {Srivani}, {Subrahmanyan}, {Sullivan}, {Tegmark},
  {Tingay}, {Trott}, {Waterson}, {Wayth}, {Webster}, {Whitney}, {Williams},
  {Williams}, {Wu}, \& {Wyithe}}]{Thyagarajan_et_al2015a}
---. 2015{\natexlab{b}}, \apj, 804, 14

\bibitem[{Tingay {et~al.}(2013)Tingay, Goeke, Bowman, Emrich, Ord, Mitchell,
  Morales, Booler, Crosse, Wayth, \& et~al.}]{Tingay2013}
Tingay, S.~J., {et~al.} 2013, \pasa, 30

\bibitem[{{Trott}(2014)}]{CTrott}
{Trott}, C.~M. 2014, \pasa, 31, e026

\bibitem[{{Trott} {et~al.}(2012){Trott}, {Wayth}, \& {Tingay}}]{Trott2012}
{Trott}, C.~M., {Wayth}, R.~B., \& {Tingay}, S.~J. 2012, \apj, 757, 101

\bibitem[{{van Haarlem, M. P.} {et~al.}(2013){van Haarlem, M. P.}, {Wise, M.
  W.}, {Gunst, A. W.}, {Heald, G.}, {McKean, J. P.}, {Hessels, J. W. T.}, {de
  Bruyn, A. G.}, {Nijboer, R.}, {Swinbank, J.}, {Fallows, R.}, {Brentjens, M.},
  {Nelles, A.}, {Beck, R.}, {Falcke, H.}, {Fender, R.}, {Hörandel, J.},
  {Koopmans, L. V. E.}, {Mann, G.}, {Miley, G.}, {Röttgering, H.}, {Stappers,
  B. W.}, {Wijers, R. A. M. J.}, {Zaroubi, S.}, {van den Akker, M.}, {Alexov,
  A.}, {Anderson, J.}, {Anderson, K.}, {van Ardenne, A.}, {Arts, M.}, {Asgekar,
  A.}, {Avruch, I. M.}, {Batejat, F.}, {Bähren, L.}, {Bell, M. E.}, {Bell, M.
  R.}, {van Bemmel, I.}, {Bennema, P.}, {Bentum, M. J.}, {Bernardi, G.}, {Best,
  P.}, {Bîrzan, L.}, {Bonafede, A.}, {Boonstra, A.-J.}, {Braun, R.}, {Bregman,
  J.}, {Breitling, F.}, {van de Brink, R. H.}, {Broderick, J.}, {Broekema, P.
  C.}, {Brouw, W. N.}, {Brüggen, M.}, {Butcher, H. R.}, {van Cappellen, W.},
  {Ciardi, B.}, {Coenen, T.}, {Conway, J.}, {Coolen, A.}, {Corstanje, A.},
  {Damstra, S.}, {Davies, O.}, {Deller, A. T.}, {Dettmar, R.-J.}, {van Diepen,
  G.}, {Dijkstra, K.}, {Donker, P.}, {Doorduin, A.}, {Dromer, J.}, {Drost, M.},
  {van Duin, A.}, {Eislöffel, J.}, {van Enst, J.}, {Ferrari, C.}, {Frieswijk,
  W.}, {Gankema, H.}, {Garrett, M. A.}, {de Gasperin, F.}, {Gerbers, M.}, {de
  Geus, E.}, {Grießmeier, J.-M.}, {Grit, T.}, {Gruppen, P.}, {Hamaker, J. P.},
  {Hassall, T.}, {Hoeft, M.}, {Holties, H. A.}, {Horneffer, A.}, {van der
  Horst, A.}, {van Houwelingen, A.}, {Huijgen, A.}, {Iacobelli, M.}, {Intema,
  H.}, {Jackson, N.}, {Jelic, V.}, {de Jong, A.}, {Juette, E.}, {Kant, D.},
  {Karastergiou, A.}, {Koers, A.}, {Kollen, H.}, {Kondratiev, V. I.},
  {Kooistra, E.}, {Koopman, Y.}, {Koster, A.}, {Kuniyoshi, M.}, {Kramer, M.},
  {Kuper, G.}, {Lambropoulos, P.}, {Law, C.}, {van Leeuwen, J.}, {Lemaitre,
  J.}, {Loose, M.}, {Maat, P.}, {Macario, G.}, {Markoff, S.}, {Masters, J.},
  {McFadden, R. A.}, {McKay-Bukowski, D.}, {Meijering, H.}, {Meulman, H.},
  {Mevius, M.}, {Middelberg, E.}, {Millenaar, R.}, {Miller-Jones, J. C. A.},
  {Mohan, R. N.}, {Mol, J. D.}, {Morawietz, J.}, {Morganti, R.}, {Mulcahy, D.
  D.}, {Mulder, E.}, {Munk, H.}, {Nieuwenhuis, L.}, {van Nieuwpoort, R.},
  {Noordam, J. E.}, {Norden, M.}, {Noutsos, A.}, {Offringa, A. R.}, {Olofsson,
  H.}, {Omar, A.}, {Orrú, E.}, {Overeem, R.}, {Paas, H.}, {Pandey-Pommier,
  M.}, {Pandey, V. N.}, {Pizzo, R.}, {Polatidis, A.}, {Rafferty, D.},
  {Rawlings, S.}, {Reich, W.}, {de Reijer, J.-P.}, {Reitsma, J.}, {Renting, G.
  A.}, {Riemers, P.}, {Rol, E.}, {Romein, J. W.}, {Roosjen, J.}, {Ruiter, M.},
  {Scaife, A.}, {van der Schaaf, K.}, {Scheers, B.}, {Schellart, P.},
  {Schoenmakers, A.}, {Schoonderbeek, G.}, {Serylak, M.}, {Shulevski, A.},
  {Sluman, J.}, {Smirnov, O.}, {Sobey, C.}, {Spreeuw, H.}, {Steinmetz, M.},
  {Sterks, C. G. M.}, {Stiepel, H.-J.}, {Stuurwold, K.}, {Tagger, M.}, {Tang,
  Y.}, {Tasse, C.}, {Thomas, I.}, {Thoudam, S.}, {Toribio, M. C.}, {van der
  Tol, B.}, {Usov, O.}, {van Veelen, M.}, {van der Veen, A.-J.}, {ter Veen,
  S.}, {Verbiest, J. P. W.}, {Vermeulen, R.}, {Vermaas, N.}, {Vocks, C.},
  {Vogt, C.}, {de Vos, M.}, {van der Wal, E.}, {van Weeren, R.}, {Weggemans,
  H.}, {Weltevrede, P.}, {White, S.}, {Wijnholds, S. J.}, {Wilhelmsson, T.},
  {Wucknitz, O.}, {Yatawatta, S.}, {Zarka, P.}, {Zensus, A.}, \& {van Zwieten,
  J.}}]{LOFAR}
{van Haarlem, M. P.} {et~al.} 2013, A\&A, 556, A2

\bibitem[{{Vedantham} {et~al.}(2012){Vedantham}, {Udaya Shankar}, \&
  {Subrahmanyan}}]{Vedantham2012}
{Vedantham}, H., {Udaya Shankar}, N., \& {Subrahmanyan}, R. 2012, \apj, 745,
  176

\bibitem[{{Zahn} {et~al.}(2012){Zahn}, {Reichardt}, {Shaw}, {Lidz}, {Aird},
  {Benson}, {Bleem}, {Carlstrom}, {Chang}, {Cho}, {Crawford}, {Crites}, {de
  Haan}, {Dobbs}, {Dor{\'e}}, {Dudley}, {George}, {Halverson}, {Holder},
  {Holzapfel}, {Hoover}, {Hou}, {Hrubes}, {Joy}, {Keisler}, {Knox}, {Lee},
  {Leitch}, {Lueker}, {Luong-Van}, {McMahon}, {Mehl}, {Meyer}, {Millea},
  {Mohr}, {Montroy}, {Natoli}, {Padin}, {Plagge}, {Pryke}, {Ruhl}, {Schaffer},
  {Shirokoff}, {Spieler}, {Staniszewski}, {Stark}, {Story}, {van Engelen},
  {Vanderlinde}, {Vieira}, \& {Williamson}}]{ZahnKSZ}
{Zahn}, O., {et~al.} 2012, \apj, 756, 65

\bibitem[{{Zheng} {et~al.}(2017){Zheng}, {Wang}, {Rhoads}, {Infante},
  {Malhotra}, {Hu}, {Walker}, {Jiang}, {Jiang}, {Hibon}, {Gonzalez}, {Kong},
  {Zheng}, {Galaz}, \& {Barrientos}}]{Rhoads2017}
{Zheng}, Z.-Y., {et~al.} 2017, \apjl, 842, L22

\end{thebibliography}

\end{document}